\documentstyle[12pt,epsf]{article}
\newcommand{\LL}{\Lambda\Lambda}

\newcommand{\gsl}{g_{\sigma \Lambda\Lambda}}
\newcommand{\gsn}{g_{\sigma NN}}
\newcommand{\gwl}{g_{\omega \Lambda\Lambda}}
\newcommand{\gwn}{g_{\omega NN}}
\newcommand{\fwl}{f_{\omega \Lambda\Lambda}}
\newcommand{\sll}{\sigma \Lambda\Lambda}
\newcommand{\wll}{\omega \Lambda\Lambda}
\newcommand{\ls}{\Lambda_{\sigma\Lambda\Lambda}}
\newcommand{\lw}{\Lambda_{\omega\Lambda\Lambda}}
\newcommand{\lf}{\Lambda_{\phi\Lambda\Lambda}}

\def\tstrut{\vrule height2.5ex depth0pt width0pt} 
\def\jtstrut{\vrule height5ex depth0pt width0pt} 
\newcommand{\be}{\begin{equation}}
\newcommand{\bea}{\begin{eqnarray}}
\newcommand{\ee}{\end{equation}}
\newcommand{\eea}{\end{eqnarray}}

\begin{document}
\begin{titlepage}
\begin{flushright}
TUM/T39-98-3\\
UG-DFM-10/97 \\
\end{flushright}  
  \vspace*{5mm}

\vspace*{2cm}

\begin{center}
{\Large \bf Double$-\Lambda$ Hypernuclei and the Nuclear Medium
Effective $\LL$ Interaction.}

\vspace{1.5cm}
{\large{\bf Jos\'e Caro$^1$, Carmen Garc\'\i a-Recio$^2$ and Juan 
Nieves$^2$}}\\[2em]
$^1$ Physik Department, Technische Universit\"at-M\"unchen,
D-85747-Garching, Germany.\\
$^2$ Departamento de F\'{\i}sica Moderna, Universidad de Granada, 
E-18071 Granada, Spain.

\end{center}

\vspace{2cm}
\begin{abstract}

We fit the  $^1 S_0$ $\LL$  interaction in the nuclear medium 
to the masses of the experimentally known double$-\Lambda$
hypernuclei: $^{\phantom{6}6}_{\LL}$He, $^{10}_{\LL}$Be and
$^{13}_{\LL}$B. We derive this effective interaction  from 
OBE J\"ulich $\LL$-type potentials and using both Hartree-Fock and variational
approaches. We find that the inclusion of $\LL$ correlations in
the variational scheme leads to significant differences and a better 
understanding of the dynamical features of the system.  
We investigate the sensitivity of the binding energies and the mesonic
decay widths of the above double$-\Lambda$ hypernuclei to the 
$\omega \Lambda\Lambda $ coupling 
 and  the form factor at the $\sigma \Lambda\Lambda$ vertex. 
 We also use this effective
interaction  to predict binding energies and pionic decay widths 
of heavier double$-\Lambda$ hypernuclei, not discovered yet. 
Finally, we discard the existence of $^1 S_0$ $\LL$ bound states 
provided the  $\LL-\Xi N$ coupling can be neglected.

\vspace{1cm}

\noindent
{\it PACS: 21.80.+a,21.30.-x,21.10.Dr,21.45+v}  \\
{\it Keywords: single and double$-\Lambda$ hypernuclei, $\LL$
 interaction, Bonn potential, J\"ulich potential,
 effective interactions, two-body correlations, mesonic decay}. 
\end{abstract}

\end{titlepage}

\newpage

\setcounter{page}{1}

\section{Introduction}

In the past years a considerable amount of work has been done both in
the experimental~(\cite{Da92}-\cite{May97}) and the theoretical
~(\cite{Br77}-\cite{OR97}) aspects of the physics of 
 single and double$-\Lambda$ hypernuclei (for a general overview see
the proceedings of the most recent {\it International Conferences 
on Hypernuclear and Strange Particle Physics}~\cite{HYP}).

Up to now, the experimental community has reported the existence 
and has measured the $^1 S_0$ ground state (angular momentum and spin of
the two interacting $\Lambda$'s coupled to zero, $L=S=0$) binding energy  of 
three double$-\Lambda$ hypernuclei\footnote{There has been
some confusion about the exact nature of the event reported by
Aoki. et. al.~\cite{Ao91}. This event was identified as
$^{10}_{\LL}$Be or $^{13}_{\LL}$B, resulting, respectively,  
in a repulsive or attractive $\LL$ interaction. However, the
theoretical analysis of Yamamoto
et. al.~\cite{Ya91} and Dover et. al.~\cite{Do91} helped to discard
the  possibility of $^{10}_{\LL}$Be, and thus it is commonly accepted an
attractive nature for the $\LL$ interaction.}: 
     $^{\phantom{6}6}_{\LL}$He, $^{10}_{\LL}$Be and $^{13}_{\LL}$B  
which energies are
reported in Table~\ref{tab:bll}. Let us consider the hypernucleus
$^{A+2}_{\LL}Z$, composed of a nuclear core $^{A}Z$ and two bound
$\Lambda$ particles. The quantity $B_{\LL}$ is defined as the total
binding energy ($> 0$) 
of the double$-\Lambda$ hypernucleus, and thus it is given by
\begin{eqnarray}
B_{\LL} & = & -\left [M\left (^{A+2}_{\LL}Z \right ) - 
M\left (^{A}Z \right ) - 2 m_\Lambda \right ],
\end{eqnarray}
\noindent where $M(\cdots)$ denotes the mass of the system which 
appears inside the brackets and $m_\Lambda$ (1115.6 MeV) is the
$\Lambda$ mass. To learn about the nature of the 
$\LL$ interaction, it is usual to define the magnitude $\Delta B_{\LL}$
as
\begin{eqnarray}
\Delta B_{\LL}  & = & B_{\LL} - 2 B_\Lambda,   \label{eq:deltabll} 
\end{eqnarray}
\noindent where $B_\Lambda$ ($> 0$),  given by 
\begin{eqnarray} 
B_{\Lambda} & = & - \left [M\left (^{A+1}_{\Lambda}Z \right ) - 
M\left (^{A}Z \right ) -  m_\Lambda \right ],
\end{eqnarray}
is the binding energy of a hyperon $\Lambda$ in the 
hypernucleus $^{A+1}_{\Lambda}Z$. Neglecting saturation effects, 
$\Delta B_{\LL}$ is suppressed with respect to $B_{\LL}$ by one power of 
the nuclear mass number. 

\begin{table}
\begin{center}
\begin{tabular}{ccc}
\hline
\tstrut
\tstrut
Hypernucleus & $B_{\Lambda\Lambda}$(MeV) & $\Delta B_{\Lambda\Lambda}$(MeV)\\ 
\hline 
\tstrut
\tstrut
$^{\phantom{6}6}_{\LL}$He  \cite{Pr66}, \cite{Fr95} &
$10.9\pm 0.8$   & $4.7\pm 1.0$ \\  
\tstrut
$^{10}_{\Lambda\Lambda}$Be
\cite{Da63}, \cite{Fr95} & $17.7 \pm 0.4$ & $4.3\pm 0.4$ \\ 
\tstrut
$^{13}_{\Lambda\Lambda}$B\phantom{e} \cite{Ao91}, \cite{Fr95} & $27.5\pm 0.7$   &
$4.8\pm 0.7$ \\
\hline
\end{tabular}
\end{center}
\caption{\small  Experimental double$-\Lambda$  hypernuclei 
binding energies, $B_{\Lambda\Lambda}$,  and values for the quantity 
$\Delta B_{\Lambda\Lambda}$ defined in
Eq.~(\protect\ref{eq:deltabll}). As it is discussed with great detail
in~\protect\cite{Do91}, 
we have considered that the event reported by~\protect\cite{Ao91}
corresponds to the double$-\Lambda$ hypernucleus 
$^{13}_{\Lambda\Lambda}$B. We use the notation
$^{A+2}_{\LL}Z$.}
\label{tab:bll}
\end{table}

Data on $\Lambda$-proton 
scattering  constitute an indirect source of information about the $\LL$ 
interaction. Although the available hyperon-nucleon scattering data are
scarce, this information has been successfully used by the
Nijmegen~\cite{Ma89,Na75}, J\"ulich~\cite{Ho89,Re94} and
T\"ubingen~\cite{St88,St90} groups to determine realistic
hyperon-nucleon and thus also some pieces of the hyperon-hyperon interactions. 
The T\"ubingen model is mainly a
constituent quark model supplemented in the long and intermediate
range part by $\pi$ and $\sigma$ exchange; the latter is treated as an
$SU(3)$ singlet, with a mass of 520 MeV. On the other hand, the
Nijmegen and J\"ulich models are based on meson exchange. In both models
most of the required intermediate range attraction is provided by the 
exchange of mesons in the scalar-isoscalar  channel. 

The J\"ulich $YN$ interaction models have been constructed following
the same ideas as those used in the Bonn $NN$ potential~\cite{Bonn}. Thus, 
these models account for the intermediate range attraction by the exchange 
of a fictitious scalar-isoscalar meson, $\sigma$, with a mass of 
about 600 MeV. The $\sigma$ is not treated as a
physical particle to which the $SU(3)$ relations should be applied,
but merely as an effective description of correlated $2\pi$ and $K\bar
K$-exchange processes. However, the Nijmegen group views  the
scalar-isoscalar interaction as generated by genuine scalar meson
exchanges ($S^*$,  $\delta$, $\epsilon$ and $\kappa$). Then, the $SU(3)$
symmetry is used to relate the couplings of the above 
mesons to the nucleons and hyperons.

J\"ulich and Nijmegen models, or some inspired in them, 
have been extensively used to analyze single--$\Lambda$
hypernuclei~\cite{Bo81,YB85,MZ89,GN93,MJ94,Zh96}.

Because the direct measurement of
the $\LL$ scattering process is impractical due to the lack of
targets, the data on $\LL$ hypernuclei provide a unique method to
learn details on the $\LL$ interaction in the vacuum. Thus, since 
the first double$-\Lambda$ hypernuclei were
discovered, suggestions to extract the vacuum $\LL$ interaction 
from these systems were
made~\cite{Bo65}. Since then, many other authors have tried to shed
light on this interesting
issue~\cite{Bo84,Bo87,Ya91,Ya92,Hi93,Sc94,Ca97}. 
Near threshold ($2m_\Lambda$)\,, the $S\,({\rm strangeness}) =-2$
baryon-baryon interaction might be described in terms of only two
couple channels~\cite{Ya91}: $\LL$ and $\Xi N$. For two $\Lambda$
hyperons bound in a nuclear medium and because of Pauli-blocking, it is
reasonable to think that the
ratio of strengths of the $\LL \to  \Xi N \to \LL$ and the diagonal 
$\LL \to \LL $ (with no $\Xi N$ intermediate states) 
transitions  is suppressed respect to the free space case. This is explicitly
shown for $^{\phantom{6}6}_{\LL}$He in Ref.~\cite{Ca97}.
Thus, the data of double$-\Lambda$
hypernuclei would mainly probe the free space diagonal $\LL$ element of the 
$\LL - \Xi N$ potential. In principle, some information about 
this piece can be extracted from the
hyperon-nucleon interaction models developed by the Nijmegen,
J\"ulich and T\"ubingen groups.

 In references~\cite{Bo84,Bo87} a variational approach
and a $\alpha$-cluster decomposition of the considered 
nuclear cores ($^4$He and $^8$Be) are used. However, neither the Nijmegen
nor the J\"ulich or T\"ubingen 
models for the $\LL$ interaction are used in these
references. In Refs.~\cite{Ya91, Ya92,Hi93,Ca97}  different Nijmegen
$\LL$ interactions are considered and $G-$matrix calculations show that the
double$-\Lambda$ hypernuclei data favor the Nijmegen model D. 
Despite of the success of J\"ulich type $\Lambda N$ 
interactions to describe the
structure of single--$\Lambda$ hypernuclei physics~\cite{MZ89,MJ94,Zh96}, 
double$-\Lambda$ hypernuclei 
studies using J\"ulich $\LL$ one boson exchange  (OBE) potentials 
have not been performed yet. 

Other problems of great interest connected with the 
$\LL$ hypernuclear physics are the possible existence of the 
$S =-2$ six
quark $H$ dibaryon~\cite{Ja77,Lo86}, and the breaking of 
$SU(3)$ symmetry in baryon-baryon interactions~\cite{Do90}. Six
quarks bound states would explore new dynamical features of the 
color interactions different to those accessible by means of the 
study of {\it normal} mesons ($q\bar{q}$) and baryons ($qqq$).
The existence of $\LL$ bound states may obstruct the experimental
detection of the dibaryon $H$. On the other hand, some authors have
shown that the existence of double$-\Lambda$ hypernuclei restricts the
feasibility of a long-living $H$ dibaryon~\cite{Do86}.

The experimental community has also invested a lot of effort in the
subject, partially aimed to discover the $H$ particle (see
Refs.~\cite{Rev-exp92}-\cite{Fr95}). Both at KEK and
at BNL facilities, cascade particles ($\Xi^-$) are being 
produced by means of the
$(K^-,K^+)$ reaction and then stopped in matter to produce $S=-2$
nuclei. At KEK it is planned~\cite{Rev-exp95} 
to increase the number of stopped 
$\Xi^-$ events by roughly a factor 10 with respect to the experience of
Refs.~\cite{Ao91,Rev-exp92}, where the double$-\Lambda$ hypernuclei 
$^{13}_{\Lambda\Lambda}$B was first discovered. At BNL the reaction 
\begin{eqnarray}
\Xi^-\, ^A Z &\to & ^{A}_{\Lambda\Lambda}\left(Z-1\right) \,\,n,
\end{eqnarray}
will be used to produce light and medium 
double$-\Lambda$ hypernuclei~\cite{Fr95,May97}.

This work is the first of a series of two where we aim to 
 describe the $\LL$ interaction, both inside of the nuclear medium and in
the vacuum. In this first work we find the effective $\LL$ potential
in the medium. Elsewhere we study the nuclear medium
modifications of the $\LL$ interaction and from this study and the
effective potential obtained here we will try to extract the $\LL$
interaction in the free space, including its spin dependence~\cite{Ca98}.

In this paper, we take the J\"ulich model 
for the interaction between the
two $\Lambda$ hyperons and fit its parameters to the binding energies
of the three experimentally known double$-\Lambda$ hypernuclei. This
model does not account for $\Xi N$ intermediate states. Thus, the
$\LL$ diagonal interaction parameters fitted here, might effectively 
include some contributions from this non-elastic intermediate channel,
though we expect them to be reasonably small, as discussed above.  
We use three different approaches: perturbative, Hartree-Fock and
variational, and find that, though the first two are practically
equivalent, the inclusion of $\LL$ correlations in the variational
scheme leads to significant differences and a better understanding of
the dynamical features of the system. We also study the
sensitivity of the hypernuclear data to the cutoff masses, used in the
J\"ulich model for the $\LL$ interaction, and  also examine potentials
with dynamical breaking of the $SU(3)$ symmetry. Thus, we end up with a
whole family of potentials describing the ground state binding energy
of the three known double$-\Lambda$ hypernuclei. With the aim of trying 
to distinguish between the different potentials, we calculate the 
mesonic decay width of boron double$-\Lambda$ hypernuclei. Finally, 
we predict binding energies and
mesonic widths of heavier hypernuclei, not detected yet,  
we discuss the possible existence of $\LL$ bound states and devote a
few words to relate the free space $\LL$ interaction to that found in
this work.

The paper is organized as follows. Our approach to the double$-\Lambda$
hypernuclei is described in Sect.~\ref{sec:model}, including 
the details on the $\LL$ interaction. In Sect.~\ref{sec:mes} 
our model to the mesonic decay of double$-\Lambda$ hypernuclei is
discussed. Results are presented in Sect.~\ref{sec:res}, which is
split in different subsections: {\it ``$\Lambda$-core potentials''}, 
{\it ``$\LL$ interaction: Hartree-Fock results''}, 
{\it ``Perturbative approach''}, 
{\it ``$\LL$ interaction: variational results''}, {\it ``Contribution
of the $\phi -$exchange'' }, {\it ``Mesonic decay
and binding energies of double$-\Lambda$ hypernuclei''} and 
{\it ``Nuclear medium  
and free space $\LL$ interactions.''} Finally in Sect.~\ref{sec:concl}
we present our conclusions. In addition, there is an appendix, where
we give some of the needed matrix elements  within the variational scheme.

\section{Model for the Double$-\Lambda$ Hypernuclei}
\label{sec:model}

We approximate the double$-\Lambda$ hypernuclei by  systems composed 
by two interacting $\Lambda$'s moving in the mean field potential 
created by the nuclear cores (${\cal V}_{\Lambda A}$).  Thus, we solve
\begin{eqnarray}
\left [\sum_{i=1,2}\left (-\frac{\vec{\nabla}_i^2}{2\mu_A} + 
{\cal V}_{\Lambda A}(\vec{r_i}) \right ) +
V_{\LL}(\vec{r_1}- \vec{r_2})
-\frac{\vec{\nabla}_1\cdot\vec{\nabla}_2}{M_A} 
+ B_{\LL}\right ] {\bf
\Phi_{\LL}}(\vec{r_1},\vec{r_2})&=&0, \label{eq:sch}
\end{eqnarray}
where $M_A$ and $\mu_A$ are the nuclear core  and 
the $\Lambda$-core reduced masses respectively, 
${\bf\Phi_{\LL}}(\vec{r_1},\vec{r_2}) $ is the wave--function of the $\LL$
pair and the $\vec{\nabla}_1\cdot\vec{\nabla}_2$ piece is 
the Hughes-Eckart (HE) term~\cite{Pa91}.  The spin dependence in
the equation above is implicit in the operators and wave--function. 
The $\Lambda$-nuclear core potential, ${\cal V}_{\Lambda A}$, 
is adjusted to reproduce the binding energies of the corresponding 
single--$\Lambda$ hypernuclei and a $\sigma$-$\omega$ meson 
exchange potential is used for the $\LL$ 
interaction in the medium, $V_{\LL}$. More details will be given 
in the next subsections.

In this approximation we neglect the dynamical re-ordering effect in the
nuclear core due to the presence of the second $\Lambda$ and assume 
that both hyperons move in the same mean field as one single hyperon
does. Both, the $\LL$ interaction and this re-ordering of the nuclear core,
contribute to $\Delta B_{\LL}$. However, the latter effect is suppressed with
respect to the former by one power of the nuclear density, which is
the natural parameter in all many body quantum theory 
expansions~\cite{FW71}. Thus, we expect the nuclear core dynamical re-ordering
effects to be of the order of $\Delta B_{\LL}/A$, that is to say around
0.5 MeV for beryllium and boron and around 1 MeV for helium. However,
in view of the large binding and incompressibility of an $\alpha$
particle, it is reasonable to think that the nuclear-core distortion
effects  in helium will not be as large as the naive estimate given
above. Thus, we will assume an uncertainty of the order of 0.5 MeV in all
hypernuclei due to these effects, which is of the order 
of the experimental errors of $B_{\LL}$ reported in Table~\ref{tab:bll}. 
This lack of our model will be translated into an increase of the size
of the systematic error of our determination of the $\LL$ potential in
the medium.

To take properly into account, in the boron region,   the effects due to 
dynamical re-ordering in the nuclear core is out of
the scope of this paper. Indeed, it would likely 
require  the use of variational montecarlo techniques to solve the system
of $A+2$ interacting particles and the use of 
realistic $NN$, $\Lambda N$ and $\LL$ interactions. The only attempt
to take partially into account these effects  can be found in
Refs.~\cite{Bo84,Bo87,Bo65} for the case of
$^{10}_{\Lambda\Lambda}$Be. There, the $^8$Be nuclear 
core is described in terms of two interacting $\alpha$ particles 
and the corresponding  four-body ($\alpha\alpha\LL$) problem is
solved and it is found that nuclear core  
polarization effects change $B_{\LL}$ by 
about 0.5 MeV, in agreement with our previous estimate based in
qualitative arguments. This model, however, is difficult to use 
for any other  nuclear core that beryllium.

Note also that the HE 
term, which naturally appears in our framework, 
contributes to $\Delta B_{\LL}$ and accounts for the
variation of the core--kinetic energy due to the presence of 
the second $\Lambda$. As we will see, this piece 
improves the simultaneous description of 
an extremely light hypernucleus, as helium,  and 
not as light ones, as  beryllium or boron.

Taking into account
Fermi statistics for the system of two identical $\Lambda$'s,  the most
general wave function in the  $^1 S_0$ channel is given by
\begin{eqnarray}
{\bf\Phi_{\LL}}(\vec{r_1},\vec{r_2}) &=& K  
\left [ f(r_1,r_2,r_{12}) 
+ f(r_2,r_1,r_{12}) \right ] ~\chi^{S=0},
\end{eqnarray}
where $K$ is a normalization constant, $r_{12} = |\vec{r_1} -
\vec{r_2}|$ and $\chi^{S=0}$ represents 
the spin-singlet.

It is also  interesting to define the 
density of probability, ${\bf  P}(r_{12})$,  
of finding the two $\Lambda$ particles at relative distance
$r_{12}$. This is given by
\begin{eqnarray}
{\bf  P}(r_{12})& = & \int d^{\,3} r_1 \int d^{\,3} r_2 \,\delta \left (|\vec{r}_1 -
\vec{r}_2| - r_{12} \right ) \left
|{\bf\Phi_{\LL}}(\vec{r_1},\vec{r_2})\right|^2. \label{eq:pr12}
\end{eqnarray}

We apply both the Hartree-Fock (HF) and variational (VAR)
approximations to solve Eq.~(\ref{eq:sch}). 

In the HF  approach, the  wave function of the two $\Lambda$'s 
is given by
\begin{eqnarray}
{\bf\Phi_{\LL}^{HF}}(\vec{r_1},\vec{r_2}) &=& \phi_\Lambda (\vec{r_1})
~ \phi_\Lambda (\vec{r_2}) 
~\chi^{S=0},  \\
\phi_\Lambda (\vec{r}) &=& R_\Lambda(r)~Y_{00}(\hat{r}),
\end{eqnarray}
where $Y_{lm}(\hat{r})$ are the spherical harmonics and  
$R_\Lambda(r)$ represents the mono-particle radial 
wave function which is  obtained using a self-consistent
procedure~\cite{Pa91}. 


 In the VAR  approach we allow for a wider family of
wave--functions  ${\bf\Phi_{\LL}}(\vec{r_1},\vec{r_2}) $, including 
$r_{12}-$correlations.  
In analogy with calculations on the atomic three--body
problem~\cite{Fra66} -- \cite{Arias94}, we use a series of standard
Hylleraas type wave functions~\cite{Hy28} to expand the wave--function of
the two $\Lambda$'s system. Thus, our ansatz for the wave function is
\begin{eqnarray}
{\bf\Phi_{\LL}^{VAR}}(\vec{r_1},\vec{r_2}) &=& \left [ \sum_{abc}^\infty 
C_{abc }(r_1^a r_2^b + r_1^b r_2^a)\, r_{12}^c \exp{\{- \alpha (r_1+r_2)\}}
\right ]
~\chi^{S=0},\label{eq:ans}
\end{eqnarray}
with $\alpha$ a real parameter,  $a,b,c$ non--negative integer numbers 
and  $ a \le b$. Note that, even in the case
$c=0$, where no correlations of the type $r_{12}$ are included, 
this variational scheme does not reduce to the HF one because our ansatz of
Eq.~(\ref{eq:ans}) can not be always gathered as the product of a function
which depends only on the radial coordinate $r_1$ times the same
function of the radial coordinate $r_2$. 
We would like to mention that, though the expected 
value of the HE term is zero in the HF ground state,
it is not when VAR wave--functions, with explicit dependence on
$r_{12}$, are used.

For practical purposes, the latter series should be truncated and
following the findings in atomic physics~\cite{Fre84} we choose to 
fix
\begin{eqnarray}
a + b + c \le N,
\end{eqnarray}
Thus, the integer number $N$ determines the dimension, $N_{\rm EL}$, of our
basis\footnote{$N_{\rm EL}$ is given by
%
\[
N_{\rm EL} = \left\{
\begin{array}{ll}
\frac{1}{24} (2 N^3 + 15 N^2 + 34 N + 24) & N\, {\rm even} \\
\\
\frac{1}{24} (2 N^3 + 15 N^2 + 34 N + 21) & N \, {\rm odd} 
\end{array}
\right.
\]
}. The
unknown parameters, $\alpha$ and $C_{abc }$, are to be determined
to give the lowest value of the energy. The linear parameters
$C_{abc}$  are obtained by solving a generalized eigenvalue
problem~\cite{PW35}, while to find the non-linear parameter
$\alpha$ one needs to use  a standard numerical minimization
algorithm. We have performed calculations 
up to $N=10$ for which we estimate that
the uncertainties in the determination of the binding energies, 
$B_{\Lambda\Lambda}$,  are one order of magnitude smaller than their 
experimental errors. Our ansatz  for ${\bf\Phi_{\LL}^{VAR}}$ and $N=10$
consists of 161 terms. In atomic physics with such a basis one gets 
precisions of the order of $10^{-6}$,~\cite{Klei90,Arias94}, 
much better than those obtained here. That is because the ratio 
between the mass of the orbiting particles and  the mass of the 
nuclear core is much more bigger here than in the case of two
electrons bound systems. The accuracy could be improved  by allowing 
for different exponential behaviors for the radial coordinates $r_1$ and  
$r_2$ and symmetrizing the resulting wave function~\cite{Arias97}. 

In the VAR approach, each of the $\Lambda$ particles is
orbiting around the nuclear core not only in the $l$=$0$ wave, 
as in the HF case. Indeed, if the angular momenta of
the two $\Lambda$'s are coupled to give zero  total 
angular momentum ($\vec{L} = \vec{l_1} + \vec{l_2}  = \vec{0} $), 
any $\Lambda$-nuclear core orbital angular momentum is permitted.  
Of course, the only possibility is that both $\Lambda$'s
have the same orbital angular momentum, $l_1=l_2=l$,  
with respect to the nuclear core. 

Given a wave function for the two
$\Lambda$'s system ${\bf\Phi_{\LL}}(r_1,r_2,r_{12})$, the
probability, ${\cal P}_l$, of finding each of the two particles with 
angular momentum $l$ and coupled to $L=0$  is given by
\begin{eqnarray}
{\cal P}_l & = & 4 \pi^2 (2 l +1
)~\int_0^{+\infty}dr_1~r_1^2~ \int_0^{+\infty}dr_2~r_2^2~\left |
\int_{-1}^{+1}d\mu~P_{l}(\mu) {\bf\Phi_{\LL}}(r_1,r_2,r_{12})
\right |^2, \label{eq:mult}
\end{eqnarray}
where $P_l$ is the Legendre Polynomial of order $l$ and $\mu$ is the
cosine of the angle between the vectors $\vec {r}_1$ and $\vec {r}_2$,
being $r_{12}=(~r_1^2+r_2^2-2 r_1 r_2 \mu)^\frac12$.

In Refs.~\cite{Bo84} and~\cite{Bo87}, 
the problem of double$-\Lambda$ hypernuclei using
a variational approach was also addressed.  However, 
the configuration space for the trial wave function
(${\bf\Phi_{\LL}^{VAR}}$), used in these references 
to describe the $\LL$ relative motion, is much more reduced than 
the one used in this work.

With a basis of the size of the one used here, one needs to compute 
more than twenty five thousand matrix elements of the Hamiltonian,
for each value of $\alpha$. Furthermore, because of the
non-orthogonality of the chosen basis there exist large 
numerical cancelations in the linear parameters 
($C_{abc}$) minimization process and it is necessary to have analytical
expressions for the matrix elements. These can be found in the
Appendix.


\subsection{Model for the $\LL$ Interaction}\label{sec:ll-int}
Given the quantum numbers of the $\Lambda$ particle, the lightest
carriers of the strong interaction  between two $\Lambda$'s, 
in the scalar-isoscalar and vector-isoscalar channels, are the  
$\sigma\,(I=0,\,J^P=0^+)$ and $\omega\,(I=0,\,J^P=1^-)$ mesons,
respectively. As it was mentioned in the introduction, the $\sigma$ is
not treated as a physical particle, but merely
 as an effective description of correlated $2\pi$ and $K\bar K$-exchange
processes, as it is done in the context of the J\"ulich models for $YN$
interaction~\cite{Ho89,Re94}. The $\eta$ and $\eta^\prime$ 
meson exchanges lead to a 
$\LL$ potentials in the pseudoscalar-isoscalar channel which
contributions are negligible at low momentum transfers
\footnote{ This is explicitly shown in Refs.~\cite{Sc94,Na75} also 
for the Nijmegen
model D.}, and are not
considered in the context of the J\"ulich models for $YN$
interaction~\cite{Ho89,Re94}. Thus, we will not consider those in
this work. 

In the spirit of the Bonn-J\"ulich potentials, we should also study 
the exchange of the $\phi\,(I=0,\,J^P=1^-)$ meson. In Ref.~\cite{Sc94}, where
the Nijmegen model D contributions of different mesons are
studied and compared for the $^1S_0~\LL$ system, it is shown that 
the contribution next in importance to that of the 
$\sigma$ and $\omega$ mesons is the $\phi -$ exchange, 
being its contribution attractive.   By taking the so-called ``ideal''
mixing angle ($\tan \theta_v = 1/\sqrt2$), the $\phi$ meson comes out as
a pure $s \bar s$ state and thus in Refs.~\cite{Ho89,Re94} 
a vanishing $\phi NN$ coupling is 
required. This provides a relation between the
singlet and octet couplings, which determines the $\phi \LL$ couplings
in terms of the $\omega \LL$ ones.

The mass of 
the $\phi$ meson is significantly larger ($m_\phi=1019.41$ MeV) than
those of the $\sigma$ and $\omega$ mesons. Then its
contribution will be relevant at very short distances, where any
potential model based on meson exchanges suffers of some
uncertainties which are translated into the inclusion of form-factors and
uncertainties in the precise values and shapes of them. 
To explore the dynamics  at short distances one needs high
momentum transfers, thus though one might expect the $\phi
-$exchange to be relevant in scattering processes at large energies
and scattering angles, it is also reasonable to think that such
contribution is much less relevant to study the dynamics of a $\LL$ pair
below threshold. 
Furthermore, because the $\phi$ meson does not couple to nucleons
within the Bonn-J\"ulich model, its contribution was not taken into
account in their phenomenological studies of the $NN$ and $YN$
scattering data. Thus, we count with any empirical
determination neither of the $\phi NN$ nor of the $\phi \LL$
form-factors. We have checked that the importance  of the $\phi
-$exchange in the $^1S_0~\LL$ system at threshold is small and,
depends strongly on
the used form-factor, as one could expect. Taking as reference the
cutoff used for the $\omega -$exchange, we find that small
changes in the cutoff lead to big changes in the small role played by the
$\phi$ meson in double$-\Lambda$ hypernuclei.  Given that we count with
only three piece of data, that we expect the $\phi -$ exchange contribution to
be small and furthermore is not
completely determined by that of the $\omega -$ exchange and thus its
inclusion would require additional free parameters, 
we will not include the $\phi -$ exchange from the beginning. We will try
to keep our model for the effective $\LL$ interaction in the medium
reasonably simple, and thus we will construct it in terms only of 
$\sigma -$ and  $\omega -$exchanges. In any case, in
~Subsect.~\ref{sec:phi}
we will discuss qualitatively and quantitatively
the influence of the $\phi -$exchange in the results presented
through  the paper.

Apart from single meson exchanges, in principle, one should also
include the $\LL-\Xi N$ coupling, as mentioned in the
Introduction. For simplicity, we will first ignore this
possibility.  The possible contribution of this coupled channel and that
of the $\phi$ meson exchange (and that of 
heavier mesons, not explicitly mentioned in this discussion) will be in
any case accounted for in the obtained effective coupling constants of
the considered $\sigma$ and $\omega$
exchanges. We will comment on this when we try to link the  
effective $\LL$
interaction obtained in this work, from the double$-\Lambda$ 
hypernuclei data, to the interaction in the free space.

The couplings of the $\sigma$ and $\omega$ mesons to the $\Lambda$ are
described by the following interaction Lagrangians~\cite{Zh96,Ho89}:
\begin{eqnarray}
{\cal L}_{\sll} & = & \gsl
\bar{\Psi}_\Lambda (x)\phi_\sigma (x)\Psi_\Lambda (x), \label{eq:lags}
\\
& &  \nonumber \\ 
{\cal L}_{\wll} & = & 
\gwl \bar{\Psi}_\Lambda (x)\gamma_\mu\phi^\mu_\omega (x)\Psi_\Lambda (x)
+
\frac{f_{\omega\Lambda\Lambda}}{4m_N}\bar{\Psi}_\Lambda (x)\sigma_{\mu\nu}
\Psi_\Lambda (x)\left (\partial^\mu\phi^\nu_\omega (x)-
\partial^\nu\phi^\mu_\omega (x) \right ) \label{eq:lagw}
\end{eqnarray}
where we have used  $m_N=938.926$\,MeV for
the nucleon mass. $\Psi_\Lambda$, $\phi_\sigma$ and $\phi_\omega$  are
the $\Lambda$, $\sigma$ and $\omega$ fields, $\gamma^\mu$ are the
Dirac matrices  and $\sigma^{\mu\nu} = i [\gamma^\mu, \gamma^\nu]/2$.
The parameter $f_{\omega\Lambda\Lambda}$ accounts 
for the tensorial part of the $\wll$ coupling.

The $\LL$ OBE potential in configuration space is
obtained via Fourier transform from  
the $\LL$ Feynman amplitudes\footnote{A non-relativistic expansion of
the amplitudes is
performed, keeping only up to quadratic terms in momenta over baryon
masses.} obtained from the above
Lagrangians and reads~\cite{Bonn}

\begin{eqnarray}
V_{\sigma}(m_\sigma,r) &=& - \frac{\gsl^2}{4\pi}m_\sigma \left
\{ \left [1- \frac14 \left (\frac{m_\sigma}{m_\Lambda}\right )^2 
\right ]Y(m_\sigma r)\right. \nonumber \\
&+& \left.\frac{1}{4m_\Lambda^2}\left [\nabla^2Y(m_\sigma r) 
 + Y(m_\sigma r) \nabla^2~\right]
+\frac12 Z_1(m_\sigma r)\vec{L~}\cdot\vec{S~}\right \},\label{eq:llcoms}\\
&&\nonumber\\
V_{\omega}(m_\omega,r) &=& \frac{\gwl^2}{4\pi}m_\omega 
\left\{ \left [1+\frac12 \left (\frac{m_\omega}{m_\Lambda}\right )^2 
\right ]Y(m_\omega r)-\frac{3}{4m_\Lambda^2}\left [\nabla^2Y(m_\omega r) 
 + Y(m_\omega r) \nabla^2~\right]\right.\nonumber \\
&+& \left. \frac16 \left (\frac{m_\omega}{m_\Lambda}\right )^2
Y(m_\omega r)\,\vec{\sigma_1}\cdot\vec{\sigma_2}-\frac32
Z_1(m_\omega r)\vec{L~}\cdot\vec{S~}-\frac{1}{12}Z(m_\omega r)S_{12} \right
\}+\nonumber\\
&+&\frac12\frac{g_{\omega
\Lambda\Lambda}f_{\omega
\Lambda\Lambda}}{4\pi}m_\omega \frac{m_\Lambda}{m_N}
\left\{ \left (\frac{m_\omega}{m_\Lambda}\right )^2 Y(m_\omega r)+\frac23 
\left (\frac{m_\omega}{m_\Lambda}\right )^2 Y(m_\omega r)\,
\vec{\sigma_1}\cdot\vec{\sigma_2}\right.\nonumber\\
&-&\left. 4Z_1(m_\omega r)\vec{L~}\cdot\vec{S~}-\frac13 Z(m_\omega r)S_{12}
\right\}\nonumber\\
&+&\frac{f_{\omega\Lambda\Lambda}^2}{4\pi}
\left (\frac{m_\Lambda}{m_N}\right )^2 m_\omega 
\left\{ \frac16 
\left (\frac{m_\omega}{m_\Lambda}\right )^2 Y(m_\omega r)
\,\vec{\sigma_1}\cdot\vec{\sigma_2}-\frac{1}{12}Z(m_\omega r)S_{12}
\right\}, \label{eq:llcomw}\\ 
&&\nonumber\\
Y(x) &=& \frac{e^{-x}}{x},\,\,\,\,\,\nabla^2 =
\frac{1}{r}\frac{d^2}{dr^2}r -\frac{\vec{L~}^2}{r^2},
\end{eqnarray}
\noindent where $m_{\sigma \,(\omega)}$ is the $\sigma\,(\omega)$ meson mass 
for which we take 550 (782.6) MeV. The operators
$\vec{\sigma},\,\vec{S},\,\vec{L}$ and $S_{12}$ are the Pauli
matrices, total spin and angular momentum and the second order rank
spin tensor operator respectively. Finally, the functions 
$Z$, $Z_1$ can be found in~\cite{Bonn}.

The potentials should also contain  form factors describing the
extended hadron structure. Although they in general depend on all the
three four-momenta involved at the vertex, they are usually
parameterized in a simple form depending only on the four-momentum, $q$,
of the exchanged meson. Taking a monopolar form and neglecting the
energy transfer dependence at the vertex~\cite{Bonn,Ho89}, we 
use form factors of the type
\begin{eqnarray}
F_\alpha(\vec{q}~) = \frac{\Lambda^2_{\alpha\LL}-m_{\alpha}^2}
{\Lambda^2_{\alpha\LL}+\vec{q}~^2},\label{eq:ff}
\end{eqnarray}
\noindent where the $\alpha$ symbol stands for  $\sigma \,\, {\rm or }
\,\, \omega$.

The use of the given form factor at each vertex leads to the 
following extended expressions for the potentials:
\begin{eqnarray}
V_{\alpha}(r) & = & 
V_{\alpha}(m_{\alpha}, r) - 
\left\{V_{\alpha}(z, r)  -\frac{\left(\Lambda_{\alpha\LL}^2-m_{\alpha}^2
\right )}{2\Lambda_{\alpha\LL}}~ \frac{\partial V_{\alpha}(z, r)  }{\partial z}
 \right \}_{z=\Lambda_{\alpha\LL}}.
\end{eqnarray}
Furthermore, for the $^1 S_0$ channel, in which we will be  
interested in this work,
one can replace
\begin{eqnarray}
\vec{\sigma_1}\cdot \vec{\sigma_2} &\to& -3, \\
\vec{L~}\cdot\vec{S~} &\to& 0,\\ 
S_{12} &\to & 0.
\end{eqnarray}
Thus, finally and after including the above substitutions,
the $\LL$ potential reads
\begin{eqnarray}\label{eq:pot-tot}
V_{\LL}^{L=S=0}(r) = V_{\sigma}^{L=S=0}(r) + V_{\omega}^{L=S=0}(r),
\end{eqnarray}
where 
\begin{eqnarray}
V_{\sigma}^{L=S=0}(r) &=& - \frac{\gsl^2}{4\pi}m_\sigma \left
\{ {\bf \tilde Y} (\sigma , r)+  \frac{1}{2m_\Lambda^2} \left 
[\left (\vec{\nabla}{\bf \tilde Y} (\sigma , r)\right)_L\cdot\vec{\nabla} +  
 {\bf \tilde Y}(\sigma , r) \nabla^2~ \right ]
\right \},\label{eq:llcoms1}\\
&&\nonumber\\
V_{\omega}^{L=S=0}(r) &=& \frac{m_\omega}{4\pi} 
\left\{  {\bf \hat g}_{\omega \Lambda\Lambda}^2 {\bf \tilde Y}(\omega , r) 
+ \frac{(\gwl^2 - {\bf \hat g}_{\omega \Lambda\Lambda}^2)}{m_\omega^2} 
\frac{(\lw^2-m_\omega^2)^2}{2 m_\omega \lw} e^{-\lw\, r}\right.\nonumber\\
&-&\left.\frac{3\gwl^2}{2m_\Lambda^2} \left [\left (\vec{\nabla}{\bf \tilde Y}
(\omega , r)\right)_L\cdot\vec{\nabla} +  
 {\bf \tilde Y}(\omega , r) \nabla^2~ \right ]
\right\}. \label{eq:llcomw1}
\end{eqnarray}
The subindex $L$ means that the operator $\vec{\nabla}$ only acts on
the function ${\bf \tilde Y}
(\alpha , r)$. On the other 
hand, ${\bf \hat g}_{\omega \Lambda\Lambda}^2$ and ${\bf \tilde
Y}(\alpha , r)$ are given by
\begin{eqnarray}
{\bf \hat g}_{\omega \Lambda\Lambda}^2 & = &  \gwl^2 - 
\frac12 \left (\frac{m_\omega}{m_\Lambda}
\right )^2  \left (\frac{3 \gwl^2}{2} +
\frac{m_\Lambda}{m_N} \gwl \fwl 
+ \left (\frac{m_\Lambda}{m_N} \right )^2 \fwl^2\right), \\
{\bf \tilde Y}(\alpha , r) & = & Y (m_\alpha r) - 
\left\{1+\frac{r}{2\Lambda_{\alpha\LL}}
\left(\Lambda_{\alpha\LL}^2-m_{\alpha}^2
\right ) \right \} \frac{\Lambda_{\alpha\LL}}{m_\alpha} 
Y (\Lambda_{\alpha\LL} r)\label{eq:ytilda}.
\end{eqnarray}

The phenomenological Lagrangians of Eqs.~(\ref{eq:lags})~--~(\ref{eq:lagw})
and form factors of the type of Eq.~(\ref{eq:ff})  
have been used in Refs.~\cite{Ho89,Re94} to describe $\Lambda N$
elastic scattering data\footnote{In these references, it is not only studied
the $\Lambda N \to \Lambda N $ process, but all hyperon-nucleon
reaction channels measured experimentally ($\Lambda N \to \Lambda N$,
$\Sigma N \to \Lambda N^\prime$ and $\Sigma N \to \Sigma^\prime
N^\prime $) using a coupled channel formalism.}. The interactions of
Ref.~\cite{Re94} are obtained from those in Ref.~\cite{Ho89} by
neglecting their small time dependence and performing again a fit to
the same set of data. The $NN$ vertices are taken from the Bonn
potential~\cite{Bonn}. The model $\hat{A}$ of Ref.~\cite{Re94} leads
to a $\LL$ potential of the same form of that of
Eqs.~(\ref{eq:llcoms})~--~(\ref{eq:llcomw}) and in Table~{\ref{tab:wsll-para}}
we give the values of the coupling constants and cutoff masses found
in this reference. Note that in Ref.~\cite{Re94}, the strength of the 
$\wll$ coupling is fixed by means of the $SU(6)$ symmetry 
($\gwl = \frac23 g_{\omega NN}$ and $f_{\omega\Lambda\Lambda} = \frac56
f_{\omega NN} - \frac12 f_{\rho NN}$) 
and thus $\gsl$, $\ls$ and $\lw$ are the
only free parameters adjusted to the empirical data. In that reference is
also pointed out that the fit has little sensitivity to the cutoff
mass $\ls$.

\begin{table}
\begin{center}
\begin{tabular}{cccc}
\hline
\tstrut
\tstrut
Vertex & $g_\alpha/\sqrt{4\pi}$ & $f_\alpha/\sqrt{4\pi}$ &
$\Lambda_\alpha$ (GeV)\\
\hline
\tstrut
$\wll$ & 2.981 & $-$2.796 & 2 \\
$\sll$ & 2.138 & $-$ & 1 \\
\hline
\end{tabular}
\end{center}
\caption{\small Coupling constants $g_\alpha$, $f_\alpha$ and cutoff masses 
$\Lambda_\alpha$ found in Ref.~\protect\cite{Re94} for model $\hat{A}$. }
\label{tab:wsll-para}
\end{table}

In principle, in this work the free parameters 
 would also be the coupling constants of the $\sigma$
($g_{\sigma\Lambda\Lambda}$) and $\omega$ ($g_{\omega\Lambda\Lambda}$
and $f_{\omega\Lambda\Lambda}$ ) mesons 
to the hyperon $\Lambda$ and the  cutoffs 
($\Lambda_{\sigma\Lambda\Lambda}$, $\Lambda_{\omega\Lambda\Lambda}$) of the
corresponding form factors. However, due to the limited set of data
available to this study, we fix $\Lambda_{\omega\Lambda\Lambda}=2$ GeV
and $f_{\omega\Lambda\Lambda}=-2.796$, as in
Table~\ref{tab:wsll-para}. To check
the sensitivity of the hypernuclear data to
$g_{\omega\Lambda\Lambda}$ and  $\Lambda_{\sigma\Lambda\Lambda}$,
we consider different families of potentials obtained by using
different values for the ratio $g_{\omega\Lambda\Lambda}/g_{\omega
NN}$ above and below the $SU(6)$ prediction, 2/3, and taking for each
value of the former ratio, two different values (1 and 2 GeV) of the cutoff
$\Lambda_{\sigma\Lambda\Lambda}$. Thus, for each potential we are left
with just one free parameter, $g_{\sigma\Lambda\Lambda}$, 
which is obtained from a best $\chi^2$ fit to the values of  
$B_{\Lambda\Lambda}$ reported in Table~\ref{tab:bll}. 

As we will
see, there is a strong correlation between the
parameters $g_{\sigma\Lambda\Lambda}$ and
$\Lambda_{\sigma\Lambda\Lambda}$, and the quantity
$g_{\sigma\Lambda\Lambda}\times 
\left (1-m^2_\sigma/\Lambda^2_{\sigma\Lambda\Lambda} \right ) $
remains constant within approximately $10\%$. This is due to the fact that
for bound $\Lambda$ particles  in the double$-\Lambda$ hypernucleus, the {\it
average} momentum transfered in the  $\LL$ vertex is much  smaller
than the typical values ($\approx 1$ GeV) of the cutoff mass. In these
circumstances the form factor at the $\sll$ vertex  can be
approximated by
\begin{eqnarray}
F_\sigma(\vec{q}~) = 1- \frac{m_{\sigma}^2}{\Lambda^2_{\sll}} 
+ {\cal O}\left (\frac{\vec{q~}^2}{\Lambda^2_{\sll}}\right ).
\label{eq:corr}
\end{eqnarray}
Similar conclusions can be drawn in the case of 
$g_{\omega\Lambda\Lambda}$ and $\Lambda_{\omega\Lambda\Lambda}$.

Finally, we would like to mention that the $\LL$ interaction determined in
this work corresponds to an effective interaction in the nuclear
medium and it is not directly comparable with that of
Ref.~\cite{Re94}. We will come back to this point in
Subsect.~\ref{sec:medium}.

\subsection{Model for the $\Lambda$-Core Interaction}
\label{sect:l-core}

The detailed study of the $\Lambda$-nuclear core dynamics is not the
aim of this paper. However, we need to check the
sensitivity of the parameters of the $\LL$ interaction, obtained from
an overall fit to the available $B_{\Lambda\Lambda}$ data, to the used
$\Lambda$-nuclear core potential. Thus, we have considered different
$\Lambda$-nucleus potentials suggested in the literature, which have
been adjusted to reproduce the binding energies $B_\Lambda$ of the
single--$\Lambda$ hypernuclei ground states. 

In this subsection we will give details of the different $\Lambda$-core
potentials used through this work. All except for one are obtained
from models to describe the dynamics of the $\Lambda N$ system.  
To obtain a $\Lambda$-nucleus potential (${\cal V}_{\Lambda A}$)
from a $\Lambda N$ interaction ($V_{\Lambda N}$) we fold the
latter with the corresponding nuclear density of the core. 
For closed--shell nuclear cores 
and $\Lambda$ particles in the ground state ($1s^\frac12$), all the
couplings involving the total angular momentum $\vec{L~}$ or
spin couplings of the type $\vec{\sigma_1}~\cdot\vec{\sigma_2}$ or $S_{12}$  
in the $V_{\Lambda N}$ potential do not contribute, and  thus for this
particular scenario the
${\cal V}_{\Lambda A}$ potential is given by
\begin{eqnarray}
{\cal V}_{\Lambda A}(r) = \int d^3r^\prime \rho_c(|\vec{r}-\vec{r}\,^\prime\,|)
V_{\Lambda N}(r^\prime),\label{eq:fold}
\end{eqnarray}
where $\rho_c$ corresponds to the density for the centers of
nucleons which is obtained from the charge density by taking into
account the finite size of the nucleon (for details see Sect.~4 of
Ref.~\cite{NOG93}).

A considerable amount of work has been devoted to the study of the
mesonic decay of the hypernucleus $^{5}_{\Lambda}$He as a mean to
investigate the repulsive part of the $\Lambda N$ interaction. Then,
different  types of $\Lambda$-$^4$He potentials have been suggested in the
literature~\cite{Mo91, St93}. Among them we have selected two
potentials, with (YNG) and without (ORG) a repulsive part
in the elementary $\Lambda N$ interaction. These potentials are
determined by the following $\Lambda N$ interactions:
\begin{eqnarray}
V_{\Lambda N}(r)^{ORG} &=& -38.19 \, e^{-(r/1.034)^2}\,\,{\rm MeV},
\label{eq:YNG}\\
&&\nonumber\\
V_{\Lambda N}(r)^{YNG} &=& \Bigl[ 919 \, e^{-(r/0.5)^2}-
206.54 \, e^{-(r/0.9)^2}-9.62 \, e^{-(r/1.5)^2} \Bigr] \,\,{\rm MeV}.\label{eq:ORG}
\end{eqnarray}
In all cases $r$ is given in fermis. 
The parameters of the above interactions have been adjusted to
reproduce the ground state binding energy of the $^5_\Lambda$He~\cite{Mo91}.

To perform a simultaneous 
analysis of the three double$-\Lambda$ 
hypernuclei we are interested in, we have also
considered other two $\Lambda$-core potentials. 

\begin{itemize}

\item In the spirit of the previous subsection
for the $\LL$ dynamics, the first $\Lambda$-core potential (SW)  
considered here, is obtained from an effective  $\sigma -\omega$ 
exchange model to describe the $\Lambda N$ interaction given by
\begin{eqnarray}
V_{\Lambda N}^\sigma(m_\sigma, r) &=& - \frac{{\bar g}_{\sigma
\Lambda \Lambda}g_{\sigma NN}}{4\pi}m_\sigma  Y(m_\sigma r), \label{eq:lns}\\
&&\nonumber\\
V_{\Lambda N}^\omega(m_\omega, r) &=&\frac{m_\omega}{4\pi} \left\{
{\bar g}_{\omega\Lambda \Lambda} g_{\omega N N} - \frac14 \left
(\frac{m_\omega^2}{m_\Lambda m_N} \right ) \right.\nonumber\\
&\times&\left. \left ( {\bar g}_{\omega\Lambda \Lambda} g_{\omega N N} -
\frac{m_\Lambda}{m_N}{\bar g}_{\omega \Lambda \Lambda }f_{\omega N N} -
g_{\omega N N }{\bar f}_{\omega \Lambda\Lambda }\right ) \right\} Y(m_\omega
r).\label{eq:lnw}
\end{eqnarray}

The potential above is a simplified version of the more general one
described in Eqs.~(38)~--~(39) of Ref.~\cite{Ma89}. Firstly, as we will
mention below, we use monopolar type form factors instead of Gaussian
type form factors as it is done there. Secondly, we have neglected all
terms of order ${\cal O}\left 
(\left ( m_{\rm meson}/m_{\rm baryon} \right )^4\right )$ and
also all terms of order 
${\cal O}\left ( m_{\rm meson}^2 (m_\Lambda^2-m_N^2)/m_{\rm baryon}^4
\right )$. Thirdly, we have neglected all type of spin terms which
will not contribute for closed--shell nuclear cores 
and $\Lambda$ particles in the ground state ($1s^\frac12$). Finally,
we have neglected non-local terms of the potential which would be 
proportional to the  $\Lambda$ momentum squared over the product of
the $\Lambda$ and nucleon masses ($\cdots (\nabla^2 \phi_\Lambda
(\vec{r}))/4m_\Lambda m_N$). For a bound particle, one expects these
contributions to be quite small.

We do not consider potentials due to the exchange of strange mesons
($K$, $K^*$, $\cdots$), which though relevant to describe the 
$\Lambda N$ scattering process~\cite{Ma89,Re94}, contribute to the binding
energy only as Fock terms and thus are suppressed at least by a
$1/A$ factor. In Ref.~\cite{Br81} these Fock contributions were
evaluated and found to be small.

To take into account the finite size of the baryons involved at the
vertices, monopolar form factors are used here as well. In coordinate
space the inclusion of form factors is implemented by means of the
substitution~\cite{Bonn}
\begin{eqnarray}
V^\alpha_{\Lambda N}(r) &=& V^\alpha_{\Lambda N}(m_\alpha ,r) 
-\frac{\Lambda^2_{\alpha NN}-m_\alpha^2}{\Lambda^2_{\alpha NN}
-\Lambda^2_{\alpha \Lambda\Lambda}} 
V^\alpha_{\Lambda N}(\Lambda_{\alpha \Lambda\Lambda} ,r)
\nonumber\\ &&\nonumber\\
&+& \frac{\Lambda^2_{\alpha
\Lambda\Lambda}-m_\alpha^2}{\Lambda^2_{\alpha NN} -\Lambda^2_{\alpha
\Lambda\Lambda}} V^\alpha_{\Lambda N}(\Lambda_{\alpha NN} ,r),\label{eq:ffs}
\end{eqnarray}
\noindent where the $\alpha$ symbol stands for  $\sigma \,\, {\rm or }
\,\, \omega$. Thus, finally the $\Lambda N$ potential reads

\begin{eqnarray}
V_{\Lambda N}^{\rm SW}(r) = V^{\sigma}_{\Lambda N}(r) +
V^{\omega}_{\Lambda N}(r). \label{eq:SW}
\end{eqnarray}

In principle, the couplings $\gwl$, $\gsl$ and 
$\fwl$ introduced in Eqs.~(\ref{eq:llcoms})~--~(\ref{eq:llcomw}) should
coincide with those appearing in Eqs.~(\ref{eq:lns})~--~(\ref{eq:lnw}), 
and denoted there with an overline. However, these couplings might be
significantly different, because in both cases we are dealing with
effective interactions ($\LL$ or $\Lambda N$ ) which parameters
also account for some contributions ($\phi-$exchange, intermediate states: $\Xi
N$, $\Sigma N$, etc..) not explicitly included. 
Furthermore, these effective interactions
are affected by renormalization effects due to the nuclear medium,
which could be significantly
different for each interaction. On top of that, as we mentioned above,
some non-localities and Fock contributions have been neglected in the
treatment of the $\Lambda$-core interaction, and their effect, though
presumably small, could affect the value of the effective couplings in
Eqs.~(\ref{eq:lns})~--~(\ref{eq:lnw}). Thus, to avoid confusions with the
$\LL$ couplings,  we denote with an overline those couplings which
appear  in the $\Lambda N$ potential.

In Eqs.~(\ref{eq:lnw})~--~(\ref{eq:ffs}) we fix the $\omega \LL$ coupling parameters
to the values given above in Table~\ref{tab:wsll-para}. The $\omega NN$ and
$\sigma NN$ coupling parameters 
are taken from the Bonn potential~\cite{Bonn} and 
compiled here in Table~\ref{tab:wsnn-para}. Finally, for each
hypernucleus ($^{A+1}_{\Lambda}Z$) we fit ${\bar g}_{\sll}$ 
to reproduce the  ground state binding energy, using two values,
1 (SW1)  and 2 GeV (SW2), for the cutoff parameter 
$\Lambda_{\sigma\Lambda\Lambda}$, as we discussed in the previous
subsection.

\begin{table}
\begin{center}
\begin{tabular}{cccc}
\hline
\tstrut
\tstrut
Vertex & $g_\alpha/\sqrt{4\pi}$ & $f_\alpha/\sqrt{4\pi}$ &
$\Lambda_\alpha$ (GeV)\\
\hline
\tstrut
$\omega NN $ & 4.472 & 0 & 1.5 \\
$\sigma NN$ & 2.385 & $-$ & 1.7 \\
\hline
\end{tabular}
\end{center}
\caption{\small Coupling constants $g_\alpha$, $f_\alpha$ and cutoff masses 
$\Lambda_\alpha$ for the $NN$ vertices. Values are taken from
Ref.~\protect\cite{Bonn}. }
\label{tab:wsnn-para}
\end{table}

\item The second $\Lambda$-nuclear core potential (BOY) considered,
 is phenomenological and it is not based
on any model for the $\Lambda N$ interaction. It was suggested long
time ago for medium nuclei by A.~Bouyssy~\cite{Bo79}. 
It has only one parameter, $V_0$, which 
we adjust for each hypernucleus ($^{A+1}_{\Lambda}Z$) to 
reproduce the ground state  binding energy. The potential reads:
\begin{eqnarray}
{\cal V}_{\Lambda A}^{\rm BOY}(r) &=& \frac{V_0}{1+e^{(r-R)/a}}\,,\nonumber\\
R &=& 1.1\, {\rm A}^{\frac13} \,\, {\rm fm}, \,\,\, a =0.6  \,\, {\rm
fm}.\label{eq:BOY}
\end{eqnarray}
A similar potential was also suggested in
 Ref.~\cite{Mi88}. There, it is also shown that, despite of not
including a spin$-$orbit part, these type of potentials give also reasonable
descriptions of non ${\bf s}$-wave $\Lambda$ bound states, in
agreement with the previous findings of Ref.~\cite{Br77}. 
\end{itemize}

\section{Mesonic Decay of Double$-\Lambda$ Hypernuclei}
\label{sec:mes}

The mesonic decay has been
computed following the method exposed in Refs.~\cite{I88} and
~\cite{NO93} which uses
shell-model baryon wave functions and distorted pion waves. 
For light nuclei in the {\it p} shell, as boron, 
the low lying state structure might not be correctly represented simply by
{\it unoccupied} single--particle orbitals, and a residual interaction
of the Cohen-Kurath type~\cite{Co65} has been added.
Reviews on the subject can be found in Refs.~\cite{OR94} and \cite{OR97}. 
We compute the 
decay widths corresponding to the following processes
\begin{eqnarray}
^{A+2}_{\Lambda\Lambda}{\rm Z} & \to & 
\left(^{A+2}_{\phantom{+2}\Lambda}{\rm
Z}\right)_{b} +
\pi^0 , \label{eq:decay1}\\
^{A+2}_{\Lambda\Lambda}{\rm Z} & \to & 
\left(^{A+2}_{\phantom{+2}\Lambda}{\rm
(Z+1)}\right)_{b} +
\pi^- ,\label{eq:decay2}\\
^{A+2}_{\Lambda\Lambda}{\rm Z} & \to & 
\left(^{A+1}_{\phantom{+2}\Lambda}{\rm Z}
\right)_{gs} + n + 
\pi^0 , \label{eq:decay4}\\
^{A+2}_{\Lambda\Lambda}{\rm Z} & \to & 
\left(^{A+1}_{\phantom{+2}\Lambda}{\rm Z}
\right)_{gs} + p + 
\pi^- ,\label{eq:decay3}
\end{eqnarray}
\noindent where $b$ denotes that the remaining $\Lambda$ is in the
ground state  and the outgoing
nucleon is in an unoccupied bound state of the 
daughter hypernucleus. On the other hand $gs$ means that
the daughter hypernucleus is left on its ground state. 
In the two last reactions the outgoing nucleon
goes to the continuum.
 
We use a model in which the pionic decay is produced by
a one-body operator
\begin{equation}
\delta \widetilde{H}_{\Lambda \pi N} = -G m_\pi^{2} \left 
\{S - ( \frac{P}{m_\pi})\,
\vec{\sigma} \cdot \vec{q}\, \right \}\, \tau^{\lambda},\label{eq:decay-op}
\end{equation}
\noindent where $(G m_\pi^{2})^{2}/8 \pi = 1.945\, 10^{-15}$, 
the constants $S$ and $P$ are equal to 1.06 and 0.527 respectively
and $m_\pi$ is the pion mass (139.6 or 135.0 MeV for $\pi^-$ or
$\pi^0$). Finally, $\vec{q}$ is the momentum of 
the outgoing pion and the 
Pauli matrices $\vec{\sigma}$ and $\tau^{\lambda}$ act on the spin and
isospin Hilbert spaces respectively. The $\tau^{\lambda}$ operator in
Eq.~(\ref{eq:decay-op}) implements the $\Delta T = 1/2$ rule by means
of which the rate of $\Lambda \rightarrow \pi^{-} p$ is twice as large
as that of $\Lambda \rightarrow \pi^{0} n$.

The  vacuum $\Lambda$ decay width is readily evaluated and 
leads for proton or neutron
decay to
\begin{eqnarray}
\Gamma^{\,(\alpha)}_{free} & = & C^{\,(\alpha)} \, 
\frac{(G m_\pi^{2})^{2}}{4 \pi}
\frac{m_N\, q_{cm}}{m_{\Lambda}}\,\left\{ S^{2} +
\left(\frac{P}{m_\pi}\right)^{2} 
q^{2}_{cm}\right \}, \label{eq:free-decay}\\
q_{cm} & = & \frac{\lambda^{1/2} (m^{2}_{\Lambda}, m_N^{2},
m_\pi^{2})}
{2m_{\Lambda}}, \\
\lambda (x,y,z) &=& x^2 + y^2 + z^2 -2xy-2xz-2yz,
\end{eqnarray}
\noindent with the isospin coefficients $C^{\,(p)} = 4$  and  
$C^{\,(n)} = 2$, and $q_{cm}$ the pion 
momentum in the center of mass frame. 
One can see from Eq.~(\ref{eq:free-decay}) that the parity violating
term, $S$,  is the dominant one in the decay.

To illustrate the main ingredients entering in the decay, we reproduce below
the decay width for any of the processes of
Eqs.~(\ref{eq:decay1})~--~(\ref{eq:decay3}), in the simple shell--model
case where 
the spin--orbit splitting of the nuclear--core levels is not taken into
account:
\begin{eqnarray}
\Gamma^{\,(\alpha)} & = &   C^{\,(\alpha)} \sum_{N \neq F} 
\int \frac{d^{\,3} q}{(2 \pi)^{3}} \frac{1}{2 \omega (q)} 2 \pi \delta
(m_{\Lambda} - B_{\LL} + B_{\Lambda} 
- \omega (q) - E_{N}) (G m_\pi^{2})^{2} \nonumber\\
& \times & \left \{ S^{2} \left\vert \int d^{\,3}x_1 \left [\int d^{\,3} x_2 
{\bf\Phi_{\LL}}(\vec{x_1},\vec{x_2}) \varphi^{\ast}_{\Lambda} 
(\vec{x_2})\right ] \, \widetilde{\varphi}
_{\pi}^{(-)} ( \vec{q} , \vec{x_1} )^{\ast} \varphi^{\ast}_{N} 
(\vec{x_1}) \right \vert^{2} \right.\nonumber\\
&+& \left. \left ( \frac{P}{m_\pi} \right)^{2} 
\left\vert \int d^{\,3}x_1 \left [\int d^{\,3} x_2 
{\bf\Phi_{\LL}}(\vec{x_1},\vec{x_2}) \varphi^{\ast}_{\Lambda} 
(\vec{x_2})\right ] \, \vec{\nabla}_1\widetilde{\varphi}
_{\pi}^{(-)} ( \vec{q} , \vec{x_1} )^{\ast} \varphi^{\ast}_{N} 
(\vec{x_1}) \right \vert^{2} \right\},\label{eq:lambda-decay}
\end{eqnarray}
where $\varphi_{N}$ and $E_{N}$ are the wave function and energy of the 
outgoing nucleon in the $\Lambda$ decay, ${\bf\Phi_{\LL}}$ 
is the wave--function of the $\LL$ pair,  
$\varphi_{\Lambda}$ is the $\Lambda$ wave function 
in the daughter hypernucleus, $\omega (q)$ 
the pion energy, and the sum over $N$ runs over the 
unoccupied nuclear orbitals given by $n$ and $l$ since spin sums are already
performed.  In Eq.~(\ref{eq:lambda-decay}) 
the sums over $N$ are over proton or neutron orbitals 
according to $\alpha$. Note that if the $\LL$ interaction and the
HE term were neglected and if  $B_{\LL}$ were replaced by
$2 B_\Lambda$ in the energy conservation delta, then 
Eq.~(\ref{eq:lambda-decay}) would yield a width of twice the corresponding
one for the decay of a $\Lambda$ in a single--$\Lambda$-hypernucleus.

The corresponding expression for the decay  
width when the spin--orbit splitting of the 
nuclear--core levels is considered, can be easily deduced from
Eqs.~(6)-(15) of Ref.~\cite{I88}.

The pion wave function $(\widetilde{\varphi}_{\pi}^{(-)} (\vec{q}, x)^{\ast})$
as a block corresponds to an incoming solution of the Klein Gordon
equation,

\begin{equation}
\left [- \vec{\bigtriangledown}^{2} + m_\pi^{2} + 2 \omega(q) V_{\rm opt}
(\vec{x})\right ] 
\widetilde{\varphi}_{\pi}^{(-)} (\vec{q}, \vec{x})^{\ast} =
(\omega(q) - V_{C} (\vec{x}))^{2} \widetilde{\varphi}_{\pi}^{(-)}
(\vec{q}, \vec{x})^{\ast},
\end{equation}

\noindent
with $V_{C} (\vec{x})$ the Coulomb potential created by the nucleus considering
finite size and vacuum polarization effects 
and $V_{\rm opt} (\vec{x})$ the optical potential
which describes the $\pi$-nucleus interaction.
This
potential has been developed microscopically and it is presented in detail
in Refs.~\cite{NOG93,NOG93bis}. It contains  the ordinary lowest
order optical potential pieces constructed from the 
$s$-- and $p$--wave $\pi N$
amplitudes. In addition second order terms in both $s$-- and $p$--waves, 
responsible for pion absorption, are also considered. 
Standard corrections, as second-order Pauli re-scattering term, ATT term,
Lorentz--Lorenz effect and long and short range nuclear correlations, 
 are also taken into account. This theoretical potential
reproduces fairly well the data  of pionic atoms (binding energies and
strong absorption widths)~\cite{NOG93} and  
low energy $\pi$-nucleus scattering~\cite{NOG93bis}.

  The $\Lambda$ wave function in the daughter
hypernucleus, $\varphi_{\Lambda}$, is computed using the
$\Lambda$-core potentials described in Subsect.~\ref{sect:l-core}. 
For the nucleons we have used the following potential~\cite{Ho71}
\begin{eqnarray}
V(r) &=&  \frac{-50 \,\, {\rm MeV}}{1 + {\rm exp} [(r -
R)/a]}\label{eq:pot-noso} ,
\end{eqnarray}
\noindent
with $R=1.25 A^{1/3}$ fm, $a=0.65 $ fm,
which provides a fair reproduction of the nuclear levels for the average
energy of major shells, as well as realistic nucleon wave 
functions.

Because, the strength of the spin--orbit force is small in the boron
region (about 4 MeV), we will neglect it, and to calculate 
the mesonic decay of the $^{13}_{\LL}{\rm B}$ hypernucleus 
we will use the potential of
Eq.~(\protect\ref{eq:pot-noso}) and Cohen-Kurath 
spectroscopic factors~\cite{Co67} for the
$1p-$shell. Those are obtained from effective interactions for the
$1p-$shell deduced in Ref.~\cite{Co65} by fitting 
different nuclear energy levels. Thus, the contribution to the mesonic
width of the 
processes where the outgoing nucleon is trapped in the $1p-$shell is
computed as follows. The initial state consists of seven nucleons in
the $1p-$configuration\footnote{The $1s_\frac12-$shell is full, and it
behaves, in a very good approximation, as an inert
core~\protect{\cite{Mac60}}.} plus two $\Lambda$  hyperons coupled to
$L=0=S$, whereas the final state consists of eight nucleons in the 
$1p-$configuration plus a
$\Lambda$ hyperon and a pion. The transition operator annihilates a
$\Lambda$ hyperon and creates a nucleon in the
$1p-$configuration. To compute the nuclear matrix elements involved in
the mesonic decay, one needs 
the angular momentum--isospin reduced matrix element 
of the nucleon creation 
operator, $a^\dagger_\rho$, between fully antisymmetric states of $n$
and $n-1$ nucleons in the angular momentum--isospin 
 configuration $\rho$  and coupled respectively to $\Gamma_R$ and
$\Gamma_i$ angular momentum and isospin quantum numbers. This is given
by~\cite{BG71},
\begin{eqnarray}
\frac{\langle \rho^n \Gamma_R ||| a^\dagger_\rho ||| \rho^{n-1}
\Gamma_i \rangle\jtstrut}{\sqrt{n}\sqrt{2\Gamma_R+1}} & = & 
\langle\rho^n \Gamma_R| \} \rho^{n-1}\Gamma_i\rangle ,  
\end{eqnarray}
where the reduced matrix element is defined by 
Eq.~(A.3.a17) of Ref.~\cite{BG71} and 
$\langle...| \} ...\rangle$ are the
coefficients of fractional parentage. 
They are related to the spectroscopic factors, ${\cal S}$, by 
\begin{eqnarray}
{\cal
S} (\Gamma_R; \Gamma_i,\rho) &=& n \,\langle\rho^n \Gamma_R| \}
\rho^{n-1}\Gamma_i\rangle ^2. 
\end{eqnarray}

After a little of Racah-algebra, one finds that the averaged modulus
squared of the nuclear matrix element ($\cal M$) 
which determines the $1p-$shell
contribution to the decay processes of
Eqs.~(\ref{eq:decay1})~--~(\ref{eq:decay2}) is given by

\begin{eqnarray}
&& \overline{\sum_i}\sum_f |{\cal M}|^2 =  \nonumber\\
&& \sum_{J_R,T_R,\alpha_R,j}
\frac{2J_R+1}{(2J_i+1)(2j+1)} {\cal C}(T_i, \frac12, T_R | \tau_i,
\tau_R-\tau_i, \tau_R)^2  
{\cal S}(J_R,T_R,\alpha_R; J_i, T_i,{\rm
gs},j) \nonumber\\
&& \times \sum_{m,m_{\Lambda_s}} \left |\left \langle \Lambda\,
\frac12\, m_{\Lambda_s}|
\delta \widetilde{H}_{\Lambda \pi N} | N  j m, \frac12
(\tau_R-\tau_i) \otimes  \pi \,\vec{q\,}(\alpha_R) \right \rangle \right |^2
\label{eq:ck}
\end{eqnarray}
where $J_i=3/2, T_i = 1/2$ and $\tau_i = -1/2$ are the total angular
momentum, isospin and third isospin component quantum numbers 
of the initial nuclear core ($^{11}{\rm B}$), $J_R, T_R$ and
$\tau_R = 0$  for $\pi^-$ decay  or $-1$ for $\pi^0$  decay, 
 are the corresponding quantum numbers for the final nuclear
core.  In addition ${\rm gs}$ and $\alpha_R$ stand for the ground state
energy of the $^{11}{\rm B}$ core and the excitation energy of the 
final cores $^{12}{\rm B^*}$ or $^{12}{\rm C^*}$ 
and $\cal C$ is a Clebsch--Gordan coefficient. 
On the other hand $j,m$ and $\tau_R-\tau_i$
determines the wave function of the outgoing nucleon after the decay.
The sum over 
the ``R'' quantum numbers gives us the sum over $p-$shell 
excited states in the
final nuclear core. For $\pi^-$ ($\pi^0$) decay, we sum 
the contribution of the eighteen (eight) excited
states which spectroscopic factors and 
energies are given in Table 1\footnote{Note that in this table the 
origin of energies is that of the $^{12}C$ ground state.} 
(label: ``Stripping for target
$A=11$ $(\frac32,\frac12)$'') of Ref.~\cite{Co67}. Finally, 
$|\pi \,\vec{q\,}(\alpha_R)  \rangle$ and $|\Lambda
m_{\Lambda_s}\rangle$ stand for the pion wave
function in the continuum, being its momentum determined by energy
conservation, and the spectator $\Lambda$ wave function, being its
spatial wave-function determined by the projection

\begin{eqnarray}\label{eq:overl}
\left  \langle \varphi_\Lambda(\vec{x}_2) | 
{\rm \Phi_{\Lambda\Lambda}}(\vec{x}_1,\vec{x}_2) \right \rangle  & = &
\int d^{\,3} x_2 
{\bf\Phi_{\LL}}(\vec{x}_1,\vec{x}_2) \varphi^{\ast}_{\Lambda} 
(\vec{x}_2).\label{eq:projection}
\end{eqnarray}

The matrix element of the interaction
 $\delta \widetilde{H}_{\Lambda \pi N}$  in Eq.~(\ref{eq:ck}) is of
the same type of that given in Eqs.~(6)-(15) of Ref.~\cite{I88}.

Further details of the calculation of the mesonic decay width as the 
treatment of the outgoing nucleons in the continuum, the correct energy
balance in the reaction, the correct treatment of the quasielastic
collisions of the outgoing pion, the procedure to perform the 
$d^{\,3}q$ and $d\Omega_1$ integrations, etc., can be found in 
Ref.~\cite{NO93}.

Experimentally, what can be observed are the inclusive processes 
\begin{eqnarray}
^{A+2}_{\Lambda\Lambda}{\rm Z} & \to & X + \Lambda + 
\pi^0 , 
\label{eq:decayq1}\\
^{A+2}_{\Lambda\Lambda}{\rm Z} & \to &  X + \Lambda +
\pi^- . 
\label{eq:decayq2}
\end{eqnarray}
The main contribution to these processes is given by the exclusive
ones shown in Eqs.~(\ref{eq:decay1})~--~(\ref{eq:decay3}). 
This can be seen by 
looking at the overlap integral defined in Eq.~(\ref{eq:projection}), 
which appears in Eq.~(\ref{eq:lambda-decay}). 
If one uses HF wave functions for the $\LL$ pair and  the
pure spectator approximation, in which the monoparticle wave function
used to construct ${\bf\Phi_{\LL}^{HF}}$ coincides with that of the $\Lambda$
ground state in the daughter hypernucleus, then, due to the orthogonality
of the $\Lambda$ wave functions, only those processes where 
the non-decaying $\Lambda$ in the 
daughter hypernucleus remains in the ground state will contribute. 
Even if these approximations are not used, we have
checked that those processes included in 
Eqs.~(\ref{eq:decayq1})~--~(\ref{eq:decayq2}) and not
included in Eqs.~(\ref{eq:decay1})~--~(\ref{eq:decay3}) represent less than
2-3\% of the total. Therefore, in what follows we will approximate the
inclusive double$-\Lambda$ hypernucleus mesonic decay widths 
(to $\pi^0$ or to $\pi^-$)  by those corresponding to the processes 
specified in Eqs.~(\ref{eq:decay1})~--~(\ref{eq:decay3}). 

\section{Results}
\label{sec:res}

Before we start discussing the results concerning double$-\Lambda$
hypernuclei, we show briefly the results for the 
$\Lambda$-core potentials described in the Subsect.~\ref{sect:l-core}.

\subsection{$\Lambda$-Core Potentials}

We have studied not only the three experimentally
observed double$-\Lambda$ hypernuclei. To study the $A$ dependence 
of $\Delta B_{\LL}$ and the mesonic decay width, 
 we have also considered the nuclear core
closed--shell hypernuclei $^{42}_{\LL}$Ca, 
$^{92}_{\LL}$Zr and $^{210}_{\LL}$Pb. Unfortunately, the
 binding energies of the corresponding 
single--$\Lambda$ hypernuclei 
($^{41}_{\phantom{1}\Lambda}$Ca, $^{91}_{\phantom{1}\Lambda}$Zr 
and $^{209}_{\phantom{12}\Lambda}$Pb ) are 
not known and we have had to approximate them by the binding energies
of the closest hypernuclei known experimentally. Thus, we have used
the experimental binding energies corresponding to 
$^{40}_{\phantom{1}\Lambda}$Ca, $^{89}_{\phantom{1}\Lambda}$Y and 
$^{208}_{\phantom{12}\Lambda}$Pb. In
Table~\ref{tab:bl}   the binding energies, 
$\Lambda$-core reduced masses and details of the charge densities for all
single--$\Lambda$ hypernuclei considered in this work can be found.

\begin{table}
\begin{center}
\begin{tabular}{clccccc}
\hline
Hypernucleus & $B_\Lambda^{exp}$ [MeV] & density & $A$ & $R\, (c)\, {\rm
[fm] }$ &
$\alpha\, (a \,{\rm [fm]})$ & $\mu_A$ (MeV) \\ \hline\tstrut
$^5_\Lambda$He &     3.12$\pm$0.02~\cite{Da92},~\cite{Bo68}     & HO
&   4 & 1.358  & 0 & 858.6 \\\tstrut
$^9_\Lambda$ Be &     6.71$\pm$0.04~\cite{Da92}      & HO  &   8 &
1.77   & 0.631 & 970.4 \\ \tstrut
$^{12}_{\phantom{1}\Lambda}$ B &   11.37$\pm$0.06~\cite{Da92}     & HO  &  11 &
1.69   & 0.811 & 1006.1\\ \tstrut
$^{40}_{\phantom{1}\Lambda}$ Ca &   18.7$\pm$ 1.1~\cite{Pi91} &    &     &
&  &     \\\tstrut
$^{41}_{\phantom{1}\Lambda}$ Ca &               & 2pF &  40 & 3.51   & 0.563  &
1083.17 \\ \tstrut
$^{89}_{\phantom{1}\Lambda}$ Y &   22.0$\pm$ 0.5~\cite{Ha96} &    &     &        &   &   \\\tstrut
$^{91}_{\phantom{1}\Lambda}$ Zr &             & 2pF &  90 & 4.84   & 0.55 & 1100.95
\\ \tstrut
$^{208}_{\phantom{21}\Lambda}$ Pb & 26.5$\pm$ 0.5~\cite{Ha96}  &    &     &
& &      \\ \tstrut
$^{209}_{\phantom{12}\Lambda}$ Pb &             & 2pF & 208 & 6.624  & 0.549 & 1109.21\\
\hline
\end{tabular}\\
\end{center}
\caption{\small Binding energies, charge densities 
(\protect\cite{JVV74}), nuclear mass numbers,  
and $\Lambda$-core reduced masses   for several
single--$\Lambda$ hypernuclei used through this work. 2pF stands for a two
parameter Fermi density, $ \rho_{\rm 2pF}(r) =
\rho_0/\left(1+e^{(r-c)/a}\right )$, and HO stands for an harmonic
oscillator density, $\rho_{\rm HO}(r) = \rho_0 \left (1+\alpha\left 
( r/R \right)^2\right ) e^{-(r/R)^2}$. The density of nucleon centers,
which appears in Eq.~(\protect\ref{eq:fold}), is obtained 
from the charge density  by taking into
account the finite size of the nucleon (for details see Sect.~4 of
Ref.~\protect\cite{NOG93}).  }
\label{tab:bl}
\end{table}

\begin{figure}
\vspace{-2.cm}
\begin{center}                                                                
\leavevmode
\epsfysize = 750pt
\makebox[0cm]{\epsfbox{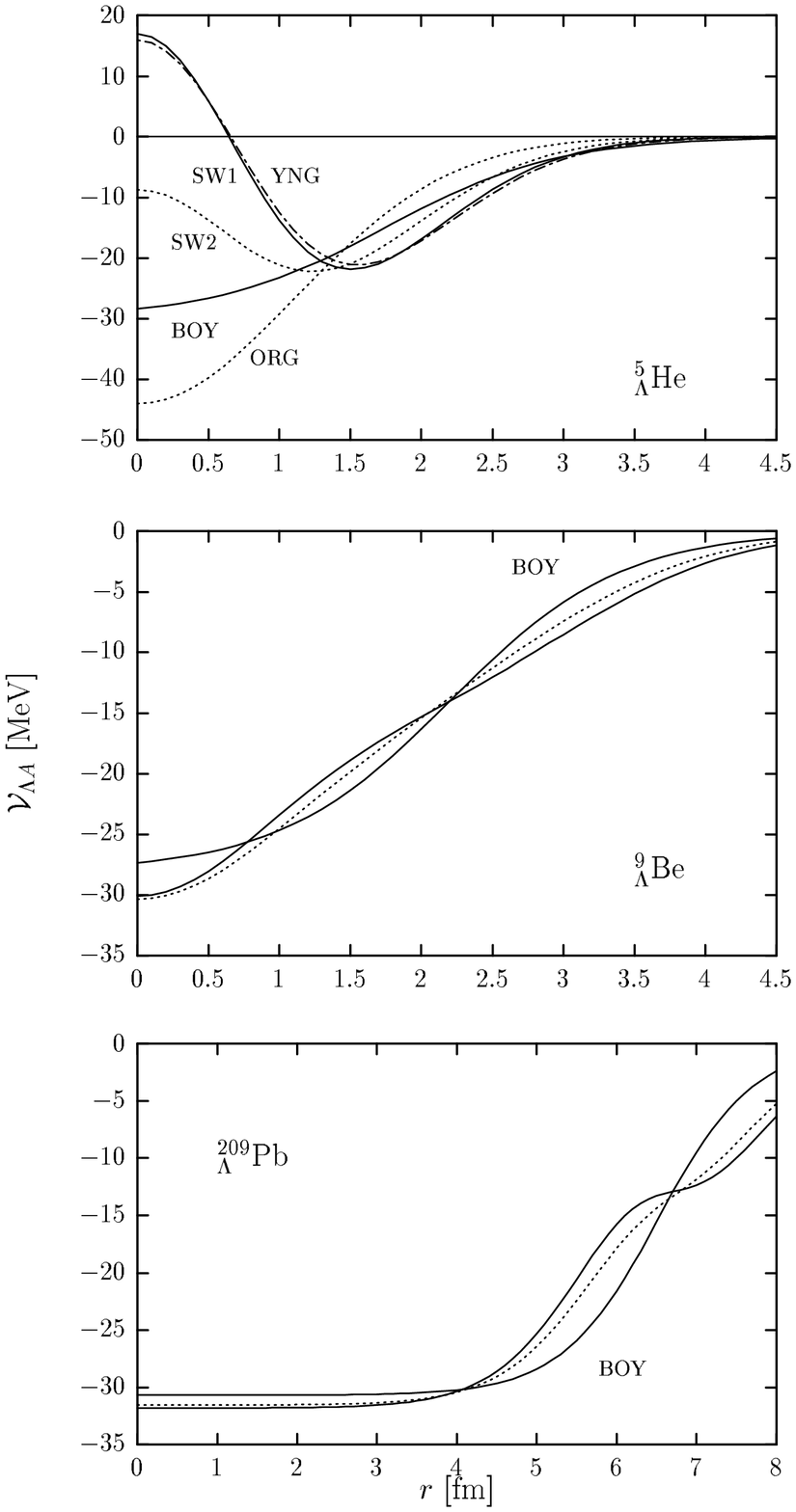}}
\end{center}
\vspace{-4.cm}
\caption{\small  $\Lambda$-core potentials (in MeV) for different hypernuclei. Solid
lines correspond always to the BOY or SW1 potentials. For
$^9_\Lambda $Be and $^{209}_\Lambda $Pb, the dotted
lines stands for the SW2 potential, and for $^5_\Lambda $He we also give
the YNG and ORG potentials.} 
\label{fig:lcore}
\end{figure}

In Table~\ref{tab:vlcore}, we show the binding energies and mean 
squared radius  obtained for the ground state  of different 
single--$\Lambda$ hypernuclei by using the $\Lambda$-core potentials 
described in Subsect.~\ref{sect:l-core}. In this table we also 
give the parameters $V_0$ and ${\bar g}_{\sigma\Lambda\Lambda}$
fitted to  the ground state binding energy for each
hypernucleus. We would like to make a few remarks: 
\begin{itemize}
\item YNG and ORG potentials were only adjusted in Ref.~\cite{Mo91} for
$^5_\Lambda$He.  Thus, we will restrict their use
only to the case of helium.
\item In the original work of Ref.~\cite{Bo79} an overall fit to
light-medium hypernuclei was performed which provided a value of
$V_0 = -32\pm 2 $ MeV, which is in good agreement with the values
shown in Table~\ref{tab:vlcore}.
\item One might think that for the case of the $\sigma -\omega$ interaction
with a cutoff $\ls=1$~GeV, the fitted parameter
${\bar g}_{\sigma\Lambda\Lambda}/\sqrt{4\pi}$ should coincide with the value
found in Ref.~\cite{Re94}, 2.138, for hyperon-nucleon scattering in the
vacuum. We find significantly different values, 3.1-3.5, which may be
due to the fact that we are dealing with a medium interaction, 
that we have neglected Fock terms and some non-localities which may be
significant for light and heavy nuclei respectively, that
we are using a $\Lambda$-core potential designed for 
closed--shell cores, for some hypernuclei where this is not the
case, and finally that we have not explicitly considered here the
channel\footnote{Though relevant far beyond the $\Lambda N$ threshold,
we expect the contribution of this new channel to be small for
$\Lambda$-bound states because of Pauli-blocking~\protect\cite{Ha93}. 
In any case it will  contribute to the above-mentioned difference
between the $\gsl$ couplings.} 
$\Lambda N \to \Sigma N$ which is included in the analysis of
Ref.~\cite{Re94}. To trace back the origin of this discrepancy is even
more complicated when one realizes that, in both this work and that of
Ref.~\cite{Re94}, for a given set of $\omega -$couplings, what is
determined is the product
of $\gsl\gsn$  rather that the  $\gsl -$ coupling alone. Hence, the
observed discrepancy could also be due to differences between the
$\sigma NN$ coupling in the vacuum and inside of the nuclear medium.

\item In contrast to the findings of Ref.~\cite{Re94}, our fits
depend appreciably on the cutoff parameter $\ls$. Indeed we find that
the quantity ${\bar g}_{\sigma\Lambda\Lambda}\times 
\left (1-m^2_\sigma/\Lambda^2_{\sigma\Lambda\Lambda} \right ) $
remains constant within approximately $2\%$, as we expected from our
discussion in Eq.~(\ref{eq:corr}). Despite of the fact that the latter
quantity remains almost constant, for very light hypernucleus as 
$^5_\Lambda$He, the short-distance behavior of the potential depends
strongly on the specific value of
the cutoff $\ls$, as can be seen in Fig.~\ref{fig:lcore}.

\item Because of the folding in Eq.~(\ref{eq:fold}), 
the  repulsive or attractive character  at short distances
of the $\Lambda N$ interaction can be appreciably
seen in the $\Lambda$-core potential only for very light hypernuclei 
as $^5_\Lambda$He. This is clearly shown in Fig.~\ref{fig:lcore},
where one can also see that for medium and heavy hypernuclei 
none of the $\Lambda$-core potentials provide repulsion at short distances 
despite of the fact that some of them are being constructed out
of $\Lambda N$ potentials with repulsive cores.

\item For $\sigma-\omega$ type
potentials, there is a strong cancelation between both components of
the interaction which is translated into a great sensitivity of the
binding energy to very small changes of the coupling constants, as the
extremely small statistical errors of the parameter ${\bar
g}_{\sigma\Lambda\Lambda}$ in Table~\ref{tab:vlcore} indicate.

\end{itemize}
\begin{table}
\begin{center}
\begin{tabular}{cclllll}\hline\tstrut

        &          & YNG   & ORG   & BOY & SW1  & SW2 \\

$\ls$ [GeV]
        &          &       &       &     & \phantom{1} 1& \phantom{1}2\\
Hypernucleus & & & & & & 
 \\\hline\tstrut
$^5_\Lambda$He 
 &$B_\Lambda$ [MeV]                
                & 3.12 & 3.10 & 3.12(2) & 3.12(2) & 3.12(2) \\
 &$\langle r^2\rangle ^{1/2}$[fm] 
                &  3.16 & 2.77  &  3.02(1) &  3.12(1) &  3.00(1) \\
 &par: $-V_0$ or $\hat{g}_{\sigma\Lambda\Lambda}$
                &       &       & 29.95(6)&  3.4916(11)&  2.5662(7)\\
\jtstrut
$^9_\Lambda $Be 
 &$B_\Lambda$ [MeV]
                &   &   & 6.71(4) & 6.71(4)  &  6.71(4) \\
 &$\langle r^2\rangle ^{1/2}$[fm] 
                &   &  &  2.46(1) &  2.61(1) &   2.533(5) \\
 &par: $-V_0$ or $\hat{g}_{\sigma\Lambda\Lambda}$
                &       &       & 28.02(7) &  3.4389(16)&   2.5436(12) \\
\jtstrut
$^{12}_{\phantom{1}\Lambda}$ B
&$B_\Lambda$ [MeV]
                & &  & 11.37(6)& 11.37(6)&11.37(6) \\
 &$\langle r^2\rangle ^{1/2}$[fm] 
                &    & &  2.216(3) &  2.259(4) & 2.183(3) \\
 &par: $-V_0$ or $\hat{g}_{\sigma\Lambda\Lambda}$
                &       &       & 31.98(10)&  3.3938 (14)$ $& 2.506(1)   \\
\jtstrut
$^{41}_{\phantom{1}\Lambda}$ Ca
&$B_\Lambda$ [MeV]
                & & & $18.7\pm 1.1$ & $18.7\pm 1.1$ &$18.7\pm 1.1 $\\
 &$\langle r^2\rangle ^{1/2}$[fm] 
                &  &  &  2.47(3) & 2.21(3) & 2.26(3) \\
 &par: $-V_0$ or $\hat{g}_{\sigma\Lambda\Lambda}$
                &       &       & $29.7\pm 1.3$&  3.194(16) & 2.390(12)   \\
\jtstrut
$^{91}_{\phantom{1}\Lambda}$ Zr
&$B_\Lambda$ [MeV]
                & & & 22.0(5)& 22.0(5)& 22.0(5)\\
 &$\langle r^2\rangle ^{1/2}$[fm] 
                &   &   &  2.91(1) &  2.562(7) & 2.645(8) \\
 &par: $-V_0$ or $\hat{g}_{\sigma\Lambda\Lambda}$
                &       &       & 28.9(5)&  3.129(7)& 2.353(5)   \\
\jtstrut
$^{209}_{\phantom{11}\Lambda}$ Pb
 &$B_\Lambda$ [MeV]
                & & & 26.5(5) & 26.5(5) & 26.5(5) \\
 &$\langle r^2\rangle ^{1/2}$[fm] 
                &   &  & 3.588(7)& 3.341(5) & 3.428(5) \\
 &par: $-V_0$ or $\hat{g}_{\sigma\Lambda\Lambda}$
                &       &        & 30.7(5)&  3.145(7)& 2.371(5) \\
\hline
\end{tabular}\\
\end{center}
\caption{\small Binding energies and $\Lambda$-mean squared radius 
obtained for the ground state of different 
single--$\Lambda$ hypernuclei by using the $\Lambda$-core potentials 
described in Eqs.~(\protect\ref{eq:YNG}), (\protect\ref{eq:ORG}), 
(\protect\ref{eq:SW}) and (\protect\ref{eq:BOY}). In the 
case of the YNG and ORG potentials , 
we use the parameters given in  Eqs.~(\protect\ref{eq:YNG}) and
(\protect\ref{eq:ORG}). For the BOY potential  we fit the
parameter $V_0$ (MeV) in Eq.~(\ref{eq:BOY}) 
whereas for the SW1 and SW2 potentials, the best fit 
parameter is $\hat{g}_{\sigma\Lambda\Lambda} = 
{\bar g}_{\sigma\Lambda\Lambda}/\protect\sqrt{4\pi}$ 
in  Eq.~(\protect\ref{eq:lns}). The errors (in brackets) in the fit
parameters and in the mean squared radius are due to the experimental
errors in the binding energies which are being fitted to.}
\label{tab:vlcore}
\end{table}

\subsection{$\LL$ Interaction: HF Results}
\label{sec:HF}

In Table~\ref{tab:HF} we show, within the HF scheme, the parameter $\gsl$
obtained from a best fit to the available $B_{\LL}$ experimental data 
(Table~\ref{tab:bll}). The value of the $\wll$ coupling is kept fixed to
$\gwl = 2 g_{\omega N N}/ 3$ and we consider several
$\Lambda$-core potentials and cutoff parameters $\ls$.
The $\chi^2$ by degree of freedom for each fit is also shown. 
YNG and ORG potentials have been considered only for helium. 

The fits using the ORG for helium and BOY for beryllium and boron 
 $\Lambda -$core interactions,  provide the best 
$\chi^2/dof $\, ($\approx 0.1$) 
among all  $\Lambda -$core interactions considered. 
Anyhow, most of the fits presented 
in Table~\ref{tab:HF} with different $\Lambda -$core interactions  
are statistically acceptable, and in all the cases 
the differences  in $\gsl$ from the different fits   are 
of the order of the statistical errors. Indeed, the fitted values 
of $\gsl$ only depends significantly on the
value of the cutoff parameter $\ls$.  Thus, the ambiguities in the
determination of the $\Lambda$-core potential do not constitute an important
obstacle to learn details about the $\LL$ interaction.

Fits using short distance repulsive $\Lambda$-helium core potentials (YNG or
SW1)   provide significantly higher values of
$\chi^2/dof$ than those obtained when the ORG or BOY potentials are
used for helium. Besides, the only statistically  
significant contribution to the $\chi^2/dof$ comes from the helium
datum. From this fact, one might conclude that short distance
repulsive potentials for helium are not favored by the
experimental value of the $^{\phantom{0}6}_{\Lambda\Lambda}$He binding
energy. That contradicts the findings of Refs.~\cite{Mo91, St93} where
it was shown that short distance repulsive $\Lambda$-helium potentials
were favored by the experimental value of the mesonic decay width
of the  $^{5}_{\Lambda}$He hypernucleus. However it is difficult
to draw any firm conclusion because, as we will see
in Subsect.~\ref{sec:VAR}, YNG and SW1 potentials for helium 
 are not so much statistically 
disfavored  when $r_{12}-$correlations are included in the $\LL$ 
wave--function. The HE term 
in Eq.~(\ref{eq:sch}), which expected value vanishes for HF-type
wave--functions but does not when $r_{12}-$correlations are included, 
improves the simultaneous description of
an extremely light hypernucleus, as helium,  and
not as light ones, as  beryllium or boron.

\begin{table}
\begin{center}
\begin{tabular}{cc|cc|c}\hline

\multicolumn{2}{c|}{$\Lambda$-core potential} &
\multicolumn{2}{c|}{$\LL$ Interaction} &  \\\tstrut 
$^4$He & $^{8}$Be - $^{11}$B & $\ls$
(GeV) & $\gsl/\sqrt{4\pi} $ & $\chi^2/dof$ \\\hline\tstrut
BOY & BOY & 1 & $3.23 \pm 0.03$ &1.2\phantom{0} \\ 
ORG & BOY & 1 & $3.22 \pm 0.03$ &0.09 \\ 
YNG & BOY & 1 & $3.23 \pm 0.03$ &3.0\phantom{0} \\ 
SW1 & SW1 & 1 & $3.27 \pm 0.03 $ &2.4\phantom{0} \\
ORG & SW1 & 1 & $3.25 \pm 0.03 $ &0.4\phantom{0} \\
YNG & SW1 & 1 & $3.27 \pm 0.03 $ &2.7\phantom{0} \\\hline 
BOY & BOY & 2 & $2.37 \pm 0.02$ &1.3\phantom{0} \\ 
ORG & BOY & 2 & $2.36 \pm 0.02$ &0.03 \\
YNG & BOY & 2 & $2.37 \pm 0.02$ &3.6\phantom{0} \\
SW2 & SW2 & 2 & $2.37 \pm 0.02 $ &2.1\phantom{0} \\
ORG & SW2 & 2 & $2.36 \pm 0.02 $ &0.6\phantom{0} \\
YNG & SW2 & 2 & $2.37 \pm 0.02 $ &4.1\phantom{0} \\\hline 
\end{tabular}
\end{center}
\caption{\small HF results for $\gsl$ obtained from a best fit 
to the $B_{\LL}$ data. The several rows correspond to 
different $\Lambda$-core potentials and different values of the cutoff
$\ls$. In all fits, the $\wll$ coupling is kept fixed to 
$\gwl = 2 g_{\omega N N}/ 3$. The last column gives the $\chi^2$ by 
degree of freedom for each fit. Errors are statistical and have been 
obtained by increasing the value of $\chi^2$ by one unit.}
\label{tab:HF}
\end{table}

 For the sake of simplicity, when quoting our final HF results, 
we take for the central value and its
statistical error the results obtained with a BOY $\Lambda -$core
interaction for all hypernuclei\footnote{The $\gsl$ values  
obtained when we use the ORG or BOY $\Lambda$-core potentials for helium 
are practically the same, and they only differ by less than one third
of the statistical error of any of them. Thus, we have decided to
quote those results obtained from a BOY $\Lambda -$helium core 
interaction in order to be able to use the
same $\Lambda -$core potential for all hypernuclei studied in this
paper.}. The 
results obtained with the rest of the 
potentials are  used to estimate  the size of the systematic error
due to the $\Lambda$-core potential uncertainties.
To be more specific, for a given value of the cutoff mass $\ls$  
we take for this systematic error 
the statistical dispersion between the different values of $\gsl$
obtained with each of the $\Lambda -$core potentials. Besides we also
include another systematic error to account for the 
dynamical re-ordering effect in the
nuclear core due to the presence of the second $\Lambda$, as we
discussed in Sect.~\ref{sec:model} after Eq.~(\ref{eq:sch}).
 
Thus for the
results of Table~\ref{tab:HF} corresponding to
$g_{\omega\Lambda\Lambda}/g_{\omega NN}=$2/3, we   
find

\begin{equation}
g_{\sigma\Lambda\Lambda}/\sqrt{4\pi}= \left \{ \begin{array}{lr} 3.23
\pm 0.03 \pm 0.02 \pm 0.03 \,\,\,\,& \ls = 1 \,\,{\rm GeV} \\   
2.37 \pm 0.02 \pm 0.01 \pm 0.02\,\,\,\,& \ls = 2 \,\,{\rm
GeV}\end{array}\right. 
\end{equation} 
where the first error is statistical and   the second  and third  
ones are  the systematic
errors due to our uncertainties in the $\Lambda-$core potential and
in our treatment of the dynamics of the nuclear core,
respectively. The latter one should account for an uncertainty of
about 0.5 MeV, as discussed in Sect.~\ref{sec:model},
 in our theoretical determination of $B_{\LL}$.  
For this systematic error we have taken an error of the
same size as the statistical one. Thus, the inclusion of this error
should account for an uncertainty of about 0.5 MeV 
(approximate  
size of the typical errors of the experimental data for $\Delta
B_{\LL}$) in our determination of the total energies of the
double$-\Lambda$ hypernuclei. By adding all errors in
quadratures, because all of them come from totally uncorrelated
sources, we finally obtain

\begin{equation}
g_{\sigma\Lambda\Lambda}/\sqrt{4\pi}= \left \{ \begin{array}{lr} 3.23
\pm 0.05  \,\,\,\,& \ls = 1 \,\,{\rm GeV} \\   
2.37 \pm 0.03 \,\,\,\,& \ls = 2 \,\,{\rm
GeV}\end{array}\right. 
\label{eq:gs-hf-values}
\end{equation} 

As
mentioned above, 
in Ref.~\cite{Re94} the hyperon-nucleon scattering data
were fitted with $g_{\omega\Lambda\Lambda}/g_{\omega NN}=$2/3,
$\Lambda_{\sigma\Lambda\Lambda}=1$ GeV and
$g_{\sigma\Lambda\Lambda}/\sqrt{4\pi }$=2.138.  This coupling constant is
not directly comparable with that in Eq.~(\ref{eq:gs-hf-values}), 
because the one determined in this work
corresponds to an effective interaction. 
We will give some more details in the
Subsects.~\ref{sec:phi} and~\ref{sec:medium}.

The $\LL$ potential used up to here has some non-localities,
namely the terms proportional to
\begin{equation}
\cdots \left [\left (\vec{\nabla}{\bf \tilde Y}
(\alpha , r)\right)_L\cdot\vec{\nabla} +  
 {\bf \tilde Y}(\alpha , r) \nabla^2~ \right ]
\label{eq:local}
\end{equation}
in Eqs.~(\ref{eq:llcoms1}) and (\ref{eq:llcomw1}). Because of the small
momentum of the $\Lambda$ particles in the hypernuclei and because
these terms are suppressed by powers of $\left 
( m_{\rm meson} / m_{\rm baryon}\right )^2$, the contribution of these
non-localities is small. In order to have a local potential to be
plotted, we have dropped out these terms and
redone the fits. We find fits of similar quality to those obtained
with the non-local $\LL$ interaction and values of $\gsl$ smaller than
the previous ones by less than about 0.5\%, indicating that the effect
of the non-local terms is negligible.

\begin{figure}
\vspace{-2cm}
\begin{center}                                                                
\leavevmode
\epsfysize = 550pt
\hspace{-6cm}\makebox[0cm]{\epsfbox{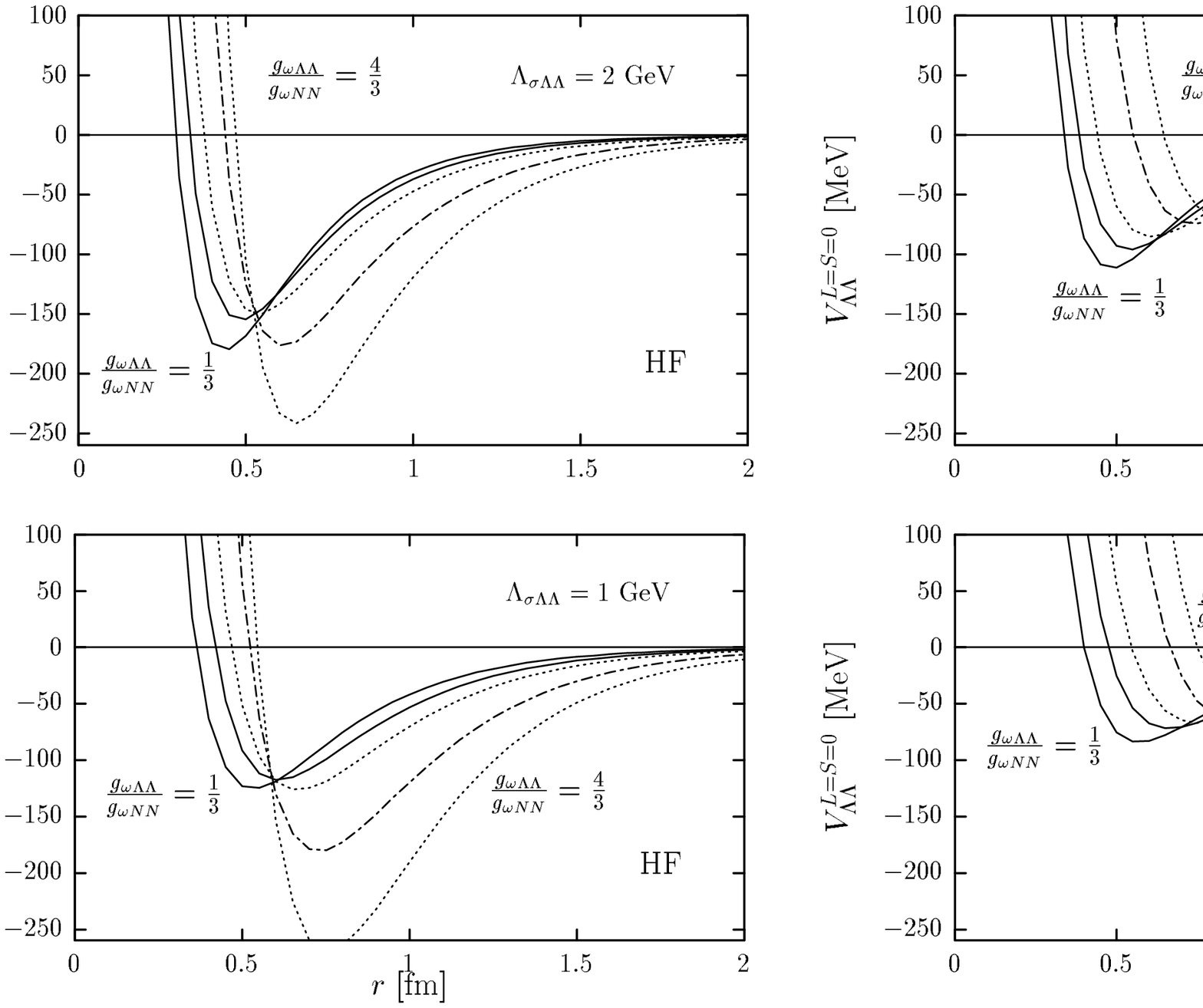}}
\end{center}
\vspace{-8.5cm}
\caption{\small Different $^1 S_0$ $\LL$ potentials, 
obtained from fits
to $B_{\Lambda\Lambda}$. Left (right) figures correspond to results
obtained within the HF (VAR) approach. In all 
figures the five curves correspond to the values 1/3, 1/2, 2/3, 1 
and 4/3 for 
the ratio $g_{\omega\Lambda\Lambda}/g_{\omega NN}$. The cutoff mass
$\Lambda_{\sigma\Lambda\Lambda}$ is 2 GeV (top) or 1  GeV
(bottom). We have used a local $\LL$ potential, as described in the 
text  after Eq.~(\protect\ref{eq:local}). BOY  $\Lambda$-core
potentials were used in all cases.}
\label{fig:hfvarll}
\end{figure}

In Fig.~\ref{fig:hfvarll}, we analyze 
the dependence of the 
$\LL$ interaction on the value of the $\wll$ coupling parameter, 
$\gwl$, and the cutoff $\ls$. 
We show different local $\LL$
potentials, obtained from best fits to the data.
Despite the different shape and magnitude of
the interactions shown in the figure, all of them give values 
for $\chi^2/dof$ of the order of 1, indicating the goodness of the
fits and the impossibility of selecting any of them only by means of the
binding energies, $B_{\Lambda\Lambda}$. The greater the parameter 
$g_{\omega\Lambda\Lambda}$, the smaller the statistical  and
systematic errors in the fitted parameter $g_{\sigma\Lambda\Lambda}$ are. 
Thus, the systematic (statistical) 
errors due to the uncertainty in the $\Lambda$-core potential 
vary from $2\%\, (2\%)$ for $g_{\omega\Lambda\Lambda}/g_{\omega
NN}=$1/3, to $0.2\%\, (0.3\%)$ for 
$g_{\omega\Lambda\Lambda}/g_{\omega NN}=$4/3. Also the $\chi^2/dof$
values decrease moderately for increasing values of the ratio $\gwl/\gwn$.

As can be seen in Fig.~\ref{fig:hfvarll} for a fixed value of the cutoff 
$\ls$, all potentials coincide at the same point. That 
implies\footnote{Let us suppose that the point where all potentials
coincide is $r_0$ and the common value of the potential is $V_0$. Then
one finds
\begin{eqnarray}
V_0 = V_{\sigma}^{L=S=0} (r_0) + V_{\omega}^{L=S=0} (r_0),\nonumber 
\end{eqnarray}
equation which automatically provides a linear relation between the 
couplings $\gsl^2$,  
 ${\bf \hat g}_{\omega \Lambda\Lambda}^2 $  
and $({\bf \hat g}_{\omega \Lambda\Lambda}^2 -
\gwl^2)$. } that there
is a linear correlation between the couplings $\gsl^2$,  
 ${\bf \hat g}_{\omega \Lambda\Lambda}^2 $ , defined in
Eq.~(\ref{eq:llcomw1}), and $({\bf \hat g}_{\omega \Lambda\Lambda}^2 -
\gwl^2)$. Indeed we find excellent fits of the couplings to a
dependence of the type
\begin{eqnarray}
\gsl^2 &=& a + b \times  {\bf \hat g}_{\omega \Lambda\Lambda}^2 + 
 c \times (\gwl^2 - {\bf \hat g}_{\omega \Lambda\Lambda}^2 ), 
\label{eq:lintotal}
\end{eqnarray}
with approximately similar values  for the parameters $b$ and $c$,
and certainly compatible within the
statistical errors. Thus, there is an accidental cancelation of the 
dependence of Eq.~(\ref{eq:lintotal}) 
on ${\bf \hat g}_{\omega \Lambda\Lambda}^2 $. To take advantage 
of this fact, we perform a two parameter fit and we find
\begin{equation}
 \frac{g_{\sigma\Lambda\Lambda}^2}{4\pi} = 
\left \{\begin{array}{lrr} (2.97 \pm 0.19) + (0.830 \pm 0.011) \times 
 \frac{\gwl^2}{4\pi}, & \chi^2/dof = 0.01, &
\Lambda_{\sigma\Lambda\Lambda} = 1 \,{\rm GeV} \\  
(1.58 \pm 0.09) + (0.448 \pm 0.005) \times
\frac{\gwl^2}{4\pi}, & \chi^2/dof = 0.02, &
\Lambda_{\sigma\Lambda\Lambda} = 2 \,{\rm GeV} \end{array}\right. 
\label{eq:swcorr-loc}
\end{equation}
for our local family of potentials.

For the the full non-local $\LL$ potential, we find

\begin{equation}
 \frac{g_{\sigma\Lambda\Lambda}^2}{4\pi} = 
\left \{\begin{array}{lrr} (3.00 \pm 0.19) + (0.836 \pm 0.011) \times 
 \frac{\gwl^2}{4\pi}, & \chi^2/dof = 0.02, &
\Lambda_{\sigma\Lambda\Lambda} = 1 \,{\rm GeV} \\  
(1.58 \pm 0.09) + (0.454 \pm 0.005) \times
\frac{\gwl^2}{4\pi}, & \chi^2/dof = 0.02, &
\Lambda_{\sigma\Lambda\Lambda} = 2 \,{\rm GeV} \end{array}\right. 
\label{eq:swcorr-noloc}
\end{equation}

To perform the fits in both the local (Eq.~(\ref{eq:swcorr-loc}))
and non-local (Eq.~(\ref{eq:swcorr-noloc})) cases, 
we add in quadrature, for each value of $\gwl$,  the
statistical and systematic errors of the fitted parameter $\gsl$.

We would like also to mention that, here again, as we expected after our
discussion in Eq.~(\ref{eq:corr}), and in
contrast to the findings of Ref.~\cite{Re94}, there is a strong 
correlation between the parameters $g_{\sigma\Lambda\Lambda}$ and
$\Lambda_{\sigma\Lambda\Lambda}$ and the quantity
$g_{\sigma\Lambda\Lambda}\times 
\left (1-m^2_\sigma/\Lambda^2_{\sigma\Lambda\Lambda} \right ) $
remains constant within approximately $3\%$. This means that, within
the HF frame of the analysis of $\LL$ hypernuclei, the use of different
cutoffs $\ls$ only amounts to redefine the $\gsl$ coupling.  Then the
point at which the potentials coincide is approximately the same for 
both sets of $\LL$ potentials  ($\ls$~=~1 GeV and $\ls$~=~2 GeV), 
as can be seen in Fig.~\ref{fig:hfvarll}.

\subsection{A Perturbative Approach}
\label{sec:pert}

The fact that the quantity $\Delta B_{\LL}/ 2 B_{\LL}$ is 
smaller than one for all hypernuclei considered in this work 
has made us to explore the possibility of using a perturbative
scheme.  Surprisingly, we find that 
in a good approximation the $\LL$
potential, $V_{\LL}$, can be treated perturbatively, as the results of
Table~\ref{tab:pert} indicate. In this table, we compare the
HF-local results of Eq.~(\ref{eq:swcorr-loc}) with those obtained from
best fits to the data using the approximation
\begin{eqnarray}
\Delta B_{\LL} = - \left \langle V_{\LL}(\vec{r_1}-
\vec{r_2})-\frac{\vec{\nabla}_1\cdot\vec{\nabla}_2}{M_A} 
\right \rangle_{\bf\Phi_{\LL}^{(0)}},\label{eq:pert}
\end{eqnarray}
where $\langle {\cal O} \rangle_{\Phi}$ denotes the expected value of
the operator ${\cal O}$ in the state $\Phi$ and $\bf\Phi_{\LL}^{(0)} =
\varphi_\Lambda(\vec{r_1}) \cdot \varphi_\Lambda(\vec{r_2}) $, where
$\varphi_\Lambda$ is the $\Lambda$-wave function in the
single--$\Lambda$ hypernucleus $^{A+1}_{\phantom{1}\Lambda}Z$. To simplify we have
also neglected the non-local terms in $V_{\LL}$. Note also that the
expected value of the HE term is zero in the state
$\bf\Phi_{\LL}^{(0)}$.

In the worst of the cases ($\gwl/\gwn = 1/3$ and $\ls =2$ GeV)  the ratio 
$\frac{\gsl^{\rm pert}-\gsl^{\rm HF}}{ \gsl^{\rm HF}} $, which accounts
for the difference between the HF and the perturbative approaches, is
only 5.9\% and its value decreases to only 0.3\% for the case 
$\gwl/\gwn = 4/3$, being this latter value of the same order as the 
the statistical error (in percentage) of the fitted parameter $\gsl^{\rm HF}$.
Perturbative values of the fitted parameter $\gsl$  are always
larger than the HF ones, though for most of the cases 
both sets of couplings are compatible within total (statistical and
systematic) errors.  The  values of $\chi^2/dof$ are about a factor
of 1.5 larger  within the perturbative approximation 
than in the HF one.

\begin{table}
\begin{center}
\begin{tabular}{c|cc|cc}\hline\tstrut
&\multicolumn{2}{c|}{$\ls = 1$ GeV}&\multicolumn{2}{c}{$\ls = 2$ GeV}\\\jtstrut
$\gwl/\gwn$ & $\frac{\gsl^{\rm pert}-\gsl^{\rm HF}}{ \gsl^{\rm HF}} $ (\%)
& stat[$\gsl^{\rm HF}$] (\%) & $\frac{\gsl^{\rm pert}-\gsl^{\rm HF}}{
\gsl^{\rm HF}} $ (\%) & stat[$\gsl^{\rm HF}$] (\%) \\[.3cm]\hline\tstrut
 1/3 & 5.2 & 1.8 & 5.9 & 1.8 \\
 1/2 & 3.2 & 1.2 & 3.7 & 1.2 \\
 2/3 & 1.9 & 0.9 & 2.5 & 0.8 \\
 1   & 0.7 & 0.5 & 1.0 & 0.5 \\
 4/3 & 0.3 & 0.3 & 0.4 & 0.3 \\\hline
\end{tabular}
\end{center}
\caption{\small Comparison of perturbative (pert) and HF results for
different types of $\LL$ interaction (the small non-local terms have
been neglected). BOY $\Lambda$-core potentials 
have been used for all hypernuclei. The columns labelled by 
stat[$\gsl^{\rm HF}$] stands for the statistical error (in percentage)
of the fitted parameter $\gsl^{\rm HF}$.}  
\label{tab:pert}
\end{table}

The fact that the perturbative and the HF schemes give the same
results within approximately $5$\% can be used to understand the existence of 
strong linear correlations of the type
of Eq.~(\ref{eq:lintotal}). For simplicity
let us discuss only the case of the local $\LL$ potential. For each
hypernucleus (helium, beryllium and boron),  because 
Eq.~(\ref{eq:pert}) is satisfied in a good approximation, one gets a
linear  relation between the couplings $\gsl^2$, $\gwl^2$ and 
${\bf \hat g}_{\omega \Lambda\Lambda}^2 $, namely
\begin{eqnarray}
\gsl^2 &\approx& \frac{ - \Delta B_{\LL}}{ 
\left \langle V_{L\sigma}^{L=S=0}/\gsl^2 
\right \rangle_{\bf\Phi_{\LL}^{(0)}}} - 
\frac{\left \langle V_{\omega}^{1}
\right \rangle_{\bf\Phi_{\LL}^{(0)}}}{ 
\left \langle V_{L\sigma}^{L=S=0}/\gsl^2 
\right \rangle_{\bf\Phi_{\LL}^{(0)}}} \times 
{\bf \hat g}_{\omega \Lambda\Lambda}^2 \nonumber\\
& -& \frac{\left \langle V_{\omega}^{2}  
\right \rangle_{\bf\Phi_{\LL}^{(0)}}}{ 
\left \langle V_{L\sigma}^{L=S=0}/\gsl^2 
\right \rangle_{\bf\Phi_{\LL}^{(0)}}} \times 
(\gwl^2 - {\bf \hat g}_{\omega \Lambda\Lambda}^2 ),\label{eq:lin}
\end{eqnarray}
where 
\begin{eqnarray}
V_{\omega}^{1}(r) & = & \frac{m_\omega}{4\pi} {\bf \tilde Y}(\omega,
r), \\
V_{\omega}^{2}(r) & = & \left [ V_{L\omega}^{L=S=0}(r) - 
 {\bf \hat g}_{\omega \Lambda\Lambda}^2 V_{\omega}^{1}(r)\right ] 
/ (\gwl^2 - {\bf \hat g}_{\omega \Lambda\Lambda}^2 ), 
\end{eqnarray}
and $V_{L\alpha}^{L=S=0}(\alpha=\sigma,\omega)$ is the local part
of $V_{\alpha}^{L=S=0}$. The  fact that we find a reasonable
simultaneous  description of the three
hypernuclei, with the same value of the parameter $\gsl$ for each value 
of $\gwl$, implies that the coefficients of the couplings in
Eq.~(\ref{eq:lin}) are rather independent of the nuclear core.

Note that in the case of the non-local potential, linear relations
between the couplings can be also obtained once the non-local terms 
of the $\LL$ potential are included in the 
right hand side of Eq.~(\ref{eq:lin}).

\subsection{$\LL$ Interaction: Variational Results}
\label{sec:VAR}

\begin{table}
\begin{center}
\begin{tabular}{cccc}\hline\tstrut
 &  & \multicolumn{2}{c}{$\gsl/\sqrt{4\pi}$} \\\tstrut
$\gwl/\gwn$ & $\ls$ (GeV) & LOCAL & NON-LOCAL \\\hline\tstrut
1/3 & 1 & $1.89\phantom{0} \pm 0.03\phantom{0} \pm
0.02\phantom{0}$ & $1.91\phantom{0} \pm 0.03\phantom{0} \pm 0.02\phantom{0}$\\
1/2 & 1 & $2.33\phantom{0} \pm 0.03\phantom{0} \pm 0.02\phantom{0}$ & $2.38\phantom{0} \pm 0.03\phantom{0} \pm 0.02\phantom{0}$\\
2/3 & 1 & $2.76\phantom{0} \pm 0.03\phantom{0} \pm
0.02\phantom{0}$ & $2.85\phantom{0} \pm 0.03\phantom{0} \pm 0.02\phantom{0}$\\
1 & 1 & $3.55\phantom{0} \pm 0.03\phantom{0} \pm
0.02\phantom{0}$ & $3.74\phantom{0} \pm 0.03\phantom{0} \pm 0.02\phantom{0}$\\
4/3 & 1 & $4.26\phantom{0} \pm 0.03\phantom{0} \pm
0.02\phantom{0}$ & $4.56\phantom{0} \pm 0.03\phantom{0} \pm
0.02\phantom{0}$\\\hline\tstrut
1/3 & 2 & $1.397 \pm 0.018 \pm 0.008$ & $1.415 \pm 0.018 \pm 0.008$\\
1/2 & 2 & $1.764 \pm 0.017 \pm 0.007$ & $1.796 \pm 0.016 \pm 0.007$\\
2/3 & 2 & $2.145 \pm 0.016 \pm 0.007$ & $2.199 \pm 0.016 \pm 0.006$\\
1 & 2 & $2.876 \pm 0.017 \pm 0.006$ & $3.005 \pm 0.017 \pm 0.006$\\
4/3 & 2 & $3.551 \pm 0.019 \pm 0.006$ & $3.772 \pm 0.018 \pm 0.006$\\\hline
\end{tabular}
\end{center}
\caption{\small VAR results for $\gsl$ obtained from best fits 
to the $B_{\LL}$ data. The several rows correspond to 
different values for the ratio $\gwl/\gwn$ and different values of the cutoff
$\ls$. Central values and their statistical uncertainties (first
set of errors) are obtained using BOY $\Lambda$-core potentials for 
all hypernuclei. The $\chi^2/dof$ values for all cases are about 
0.4-0.5. Statistical errors are obtained by
increasing the value of $\chi^2$ by one unit. 
The second set of errors account for the dispersion
in the results due to the use of different $\Lambda$-core potentials,
as it was discussed in Subsect.~\protect\ref{sec:HF}. To account for 
dynamical re-ordering effects in the
nuclear core due to the presence of the second $\Lambda$, 
further systematic errors should be added. As it was discussed in
Subsect.~\protect\ref{sec:HF}, the statistical errors
provide a reasonable estimate of these additional systematic errors.  
We give results
for the full \protect$\LL$ interaction (non-local) 
and the local reduction of it discussed 
in the text after Eq.~(\protect\ref{eq:local}).}
\label{tab:var}
\end{table}

In Table~\ref{tab:var}, for given values of the ratio
$\gwl / g_{\omega NN}$ and of the cutoff
$\Lambda_{\sigma\Lambda\Lambda}$, we show the values of the parameter $\gsl$
obtained from best fits  to the $B_{\LL}$ binding energies 
by using the variational method described in
Subsect.~\ref{sec:model}. 
We use the same procedure to give the
central values and the statistical and systematic errors as in the HF
case. In all situations (i.e., different ratios $\gwl/\gwn$, 
different values for 
the cut-off parameter $\ls$ and $\Lambda$-core potentials), we find
statistically acceptable fits.  We would like just to mention that the use of
the ORG potential for  
helium provides smaller $\chi^2/dof$  than the BOY $\Lambda$-core
potential  used to quote the central value of the fitted 
parameter. Namely, values of $\chi^2/dof$ 
smaller than 0.1 for the ORG potential 
versus values around  0.5 for the BOY case. The HE term
improves significantly the simultaneous description of all three
hypernuclei when $\Lambda$-core potentials (BOY,YNG, SW1 or SW2)
not as attractive at short distances as the ORG one are used 
for helium. To be more
specific, for the case of the BOY (YNG) potential,
 if the HE term is not included the typical values for 
$\chi^2/dof$ are of the order of 1.5 (4.0), between two or  three 
times larger than those obtained when the HE piece is considered. 
In all cases the largest 
contribution to the $\chi^2/dof$ comes from the helium piece of data.

Local and non-local potentials lead
to different values for the fitted coupling $\gsl$. These changes vary
from 1\% to 7\%  when the ratio $\gwl/\gwn$ increases and they are
significantly greater than the statistical errors (except for the lowest
value of $\gwl$) and much more appreciable than those found within the
HF scheme. Variations of the cutoff mass $\ls$ lead to 
changes in the quantity $g_{\sigma\Lambda\Lambda}\times 
\left (1-m^2_\sigma/\Lambda^2_{\sigma\Lambda\Lambda} \right ) $
from  $3\%$  to $10\%$ when the ratio $\gwl/\gwn$ increases from 
1/3 to 4/3. These changes are again much more pronounced here than in the
HF case, indicating that the variational wave functions for the two
$\Lambda$'s system lead to bigger values of the transfered momentum
between both particles. This is also consistent with the fact,
mentioned above, that within the VAR scheme non-local and local
potentials lead to differences much more significant than in the HF
one.

In Fig.~\ref{fig:hfvarll} (right) we show the local $\LL$ potentials
corresponding to the different combinations of the ratio $\gwl/\gwn$
and the cutoff mass parameter $\ls$ given in Table~\ref{tab:var}. 
In this figure, for comparison, and in the same scale, the
HF local potentials discussed in the previous section are also shown. VAR 
potentials are significantly less attractive than HF ones, as the
variational principle ensures. The larger the ratio $\gwl/\gwn$, 
the bigger this effect is. 

For both local and non-local VAR potentials, we find that the coupling
constants are correlated in the form given in Eq.~(\ref{eq:lintotal}),
but with  poorer $\chi^2/dof$ values (with  
$\chi^2/dof$ ranging from 0.12, for a local $\LL$ interaction and $\ls$ =
1~GeV, to 0.8, for a local $\LL$ interaction and $\ls$ = 2~GeV) than 
in the HF case.  But, for this variational scenario, 
the values of parameters $b$ and $c$ of Eq.~(\ref{eq:lintotal}) 
are not similar,  and a linear
relationship between $\gsl^2$ and $\gwl^2$ can not be casted. 
Therefore, unlike the HF families,  the obtained families of 
variational potentials do not coincide at a single point\footnote{ The
equation $d V_{\LL}(r)/d\gwl = 0$ has as solution a value of $r$
independent of $\gwl$ (what implies that all potentials coincide in a 
single point) only if there exist a linear relation 
between the couplings $\gsl^2$ and $\gwl^2$ and a small term
proportional to $ (m_\omega/m_\Lambda)^2 \gwl \fwl $,  
in the $\omega -$exchange part of the potential is neglected.}. 

The existence of a relation of the type of Eq.~(\ref{eq:lintotal})
also in the VAR scheme can be understood if one applies the same type
of considerations used in the discussion of 
Eqs.(\ref{eq:pert})~--~(\ref{eq:lin}) to the relation
\begin{eqnarray}
B_{\LL} = - \left \langle \sum_{i=1,2}\left (-\frac{\vec{\nabla}_i^2}{2\mu_A} + {\cal V}_{\Lambda A}(\vec{r_i}) \right ) +
V_{\LL}(\vec{r_1}- \vec{r_2})
-\frac{\vec{\nabla}_1\cdot\vec{\nabla}_2}{M_A}
\right \rangle_{\bf\Phi_{\LL}^{VAR}}\,.
\end{eqnarray}

\begin{figure}
\vspace{-1cm}
\begin{center}                                                                
\leavevmode
\epsfysize = 550pt
\hspace{-6cm}\makebox[0cm]{\epsfbox{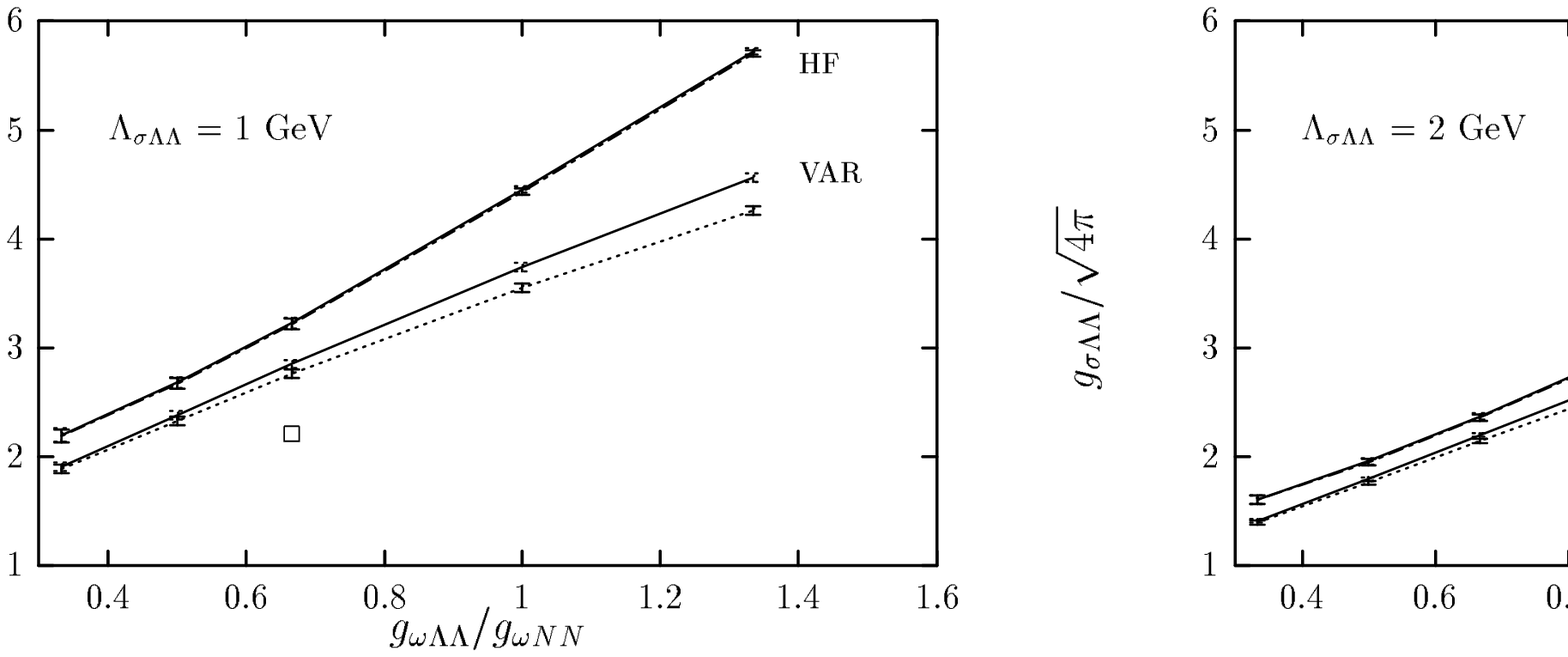}}
\end{center}
\vspace{-13.5cm}
\caption{\small Different couplings $\gsl/\protect\sqrt{4\pi}$ versus  $\gwl/\gwn$ 
found for the $^1 S_0$ $\LL$ potentials 
obtained from best fits to $B_{\Lambda\Lambda}$ data within the HF and
VAR approaches. 
The cutoff mass $\Lambda_{\sigma\Lambda\Lambda}$ is 1 GeV (left) 
or 2  GeV (right). In both plots the solid (dotted) lines
correspond to a non-local (local, as described in the 
text after Eq.~(\protect\ref{eq:local})) $\LL$ interaction and
 the five points with errors correspond to the values 1/3, 1/2, 2/3, 1 
and 4/3 for the ratio $g_{\omega\Lambda\Lambda}/g_{\omega NN}$. 
The lines have been drawn only for guiding the eye. 
BOY  $\Lambda$-core potentials 
were used in all cases. For comparison, the square in the left figure
stands for the potential of model $\hat{A}$ of
Ref.~\protect\cite{Re94}.}
\label{fig:HF-var}
\end{figure}

In Fig.~\ref{fig:HF-var} we show graphically the differences  between
the fitted parameter $\gsl$ obtained in the HF and VAR schemes for
different values of $\gwl$, different cutoffs $\ls$ and using local
(dotted lines) and non-local (solid lines) $\LL$ interactions. The
non-local part of the $\LL$ potential is repulsive, as can be deduced from 
the figure. However, the differences between VAR local 
and non-local parameters are always much smaller than the differences
between HF and VAR parameters. 

\begin{figure}
\vspace{-2.cm}
\begin{center}                                                                
\leavevmode
\epsfysize = 750pt
\makebox[0cm]{\epsfbox{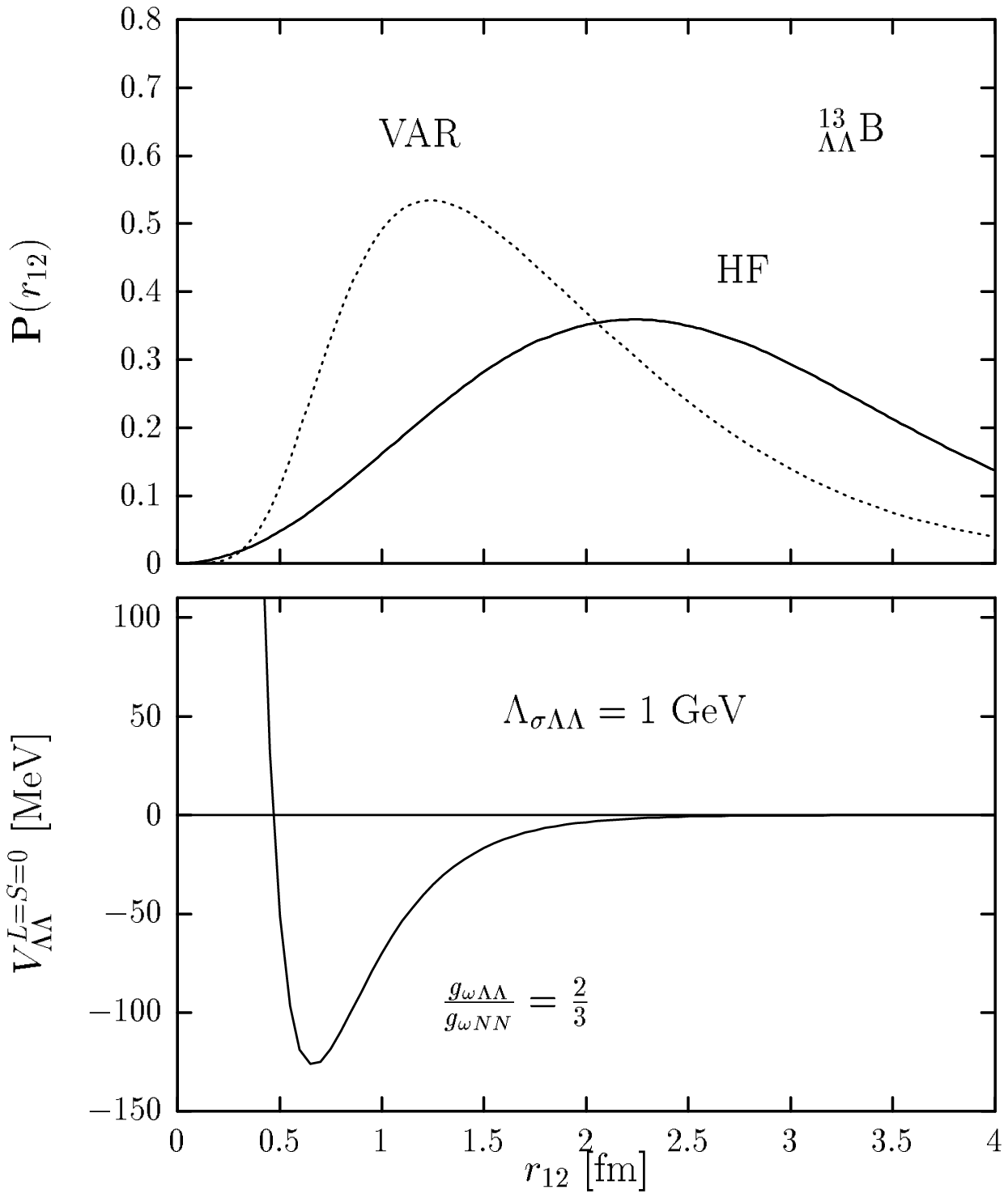}}
\end{center}
\vspace{-12cm}
\caption{\small  Top: HF (solid) and VAR (dotted) probabilities 
of finding the two $\Lambda$'s at a relative
distance $r_{12}$ as a function of this distance. Computations were
done in $^{13}_{\LL}{\rm B}$ and using the $\LL$ interaction (bottom
figure) specified in Table~\protect\ref{tab:hfvsvar}.}
\label{fig:pr12}
\end{figure}

Once we have seen that within the HF approach (or even in the
perturbative approximation) we could described
reasonably well the experimental data, one might question about the need
 of performing a variational study of the double$-\Lambda$ 
hypernuclei. To check this in
Table~\ref{tab:hfvsvar} we compare, for a fixed $\LL$ interaction, 
the results of the perturbative, HF and VAR approaches for 
the binding energy of $^{13}_{\LL}{\rm
B}$. We also show the result obtained if no correlations
(i.e., dependence on $r_{12}$) are included in the variational wave function 
(that corresponds to fix $c=0$ in the variational basis of
Eq.(\ref{eq:ans})). Even in this latter case the VAR scheme does not 
reduce to the HF one because the  wave function can not be gathered 
as a product of a function of $r_1$ times the same function now of $r_2$.
The HF approach can be considered as a variational one, where the
space of functions are limited to those which can be written in a
factorizable way. The numbers of the table show that enlarging this
space to include some non-factorizable wave--functions (but without
dependence on $r_{12}$ yet) does not modify drastically the HF results.
However, the inclusion of $r_{12}-$correlations has a 
drastic effect in the binding
energy. This can be also appreciated 
in Fig.~\ref{fig:pr12}, where
 the dependence on $r_{12}$ of the  $\LL$ potential 
(used to obtain the numbers of Table~\ref{tab:hfvsvar}) 
 and the HF and VAR densities of probability  of finding the two 
$\Lambda$'s at a relative distance $r_{12}$ (${\bf
P}(r_{12})$, defined in Eq.~(\ref{eq:pr12})) are shown. 
The effect of the inclusion of $r_{12}-$correlations is
twofold. First, it reduces the probability of finding the two
particles at very short distances, where the potential is highly
repulsive,  with values at $r_{12}=0$ as large as tens of thousands of MeV.
Second, it increases the probability of finding the particles at 
relative distances of the order of 1 fm, where the potentials have their
maximum attraction. Thus, as a net effect an important reduction 
of the expected value of the energy of the system is provided.

\begin{table}
\begin{center}
\begin{tabular}{cc}\hline\tstrut
 & $B_{\LL} \left [ ^{13}_{\LL}{\rm B} \right ]$  [MeV]\\[.2cm]\hline\tstrut 
PERT. & 27.3 \\
HF & 27.8 \\
VAR NO CORR. & 28.4 \\ 
VAR & 36.9 \\\hline
\end{tabular}
\end{center}
\caption{\small Comparison of the results obtained for 
$B_{\Lambda\Lambda}$ in boron double$-\Lambda$ hypernuclei, 
using different approximations. In all
cases the same local $\LL$ interaction (with parameters 
$\gsl/\protect\sqrt{4\pi}$, $\gwl/\gwn$ and $\ls$ fixed to
3.22, 2/3 and 1 GeV respectively) and 
BOY $\Lambda$-core potential were used.   
The coupling $\gsl$  was obtained from a best fit to the 
data within the HF approach (see results of
Eq.~(\protect\ref{eq:swcorr-loc})).   
The first, second and fourth rows correspond to the standard
perturbative, HF and VAR
calculations. The third (labeled VAR NO CORR.)  
row corresponds to a variational
calculation where the ansatz wave function does not depend on $r_{12}$
(that corresponds to fix $c=0$ in the variational basis of
Eq.(\protect\ref{eq:ans})). The experimental binding energy for boron 
double$-\Lambda$ hypernuclei is $27.5 \pm 0.7$ MeV.}
\label{tab:hfvsvar}
\end{table}

The $\LL$ interactions obtained by $\sigma - \omega$ exchange behave 
almost like a hard-core at short distances, then in a very small
region and close to the origin ($r_{12}\approx 0.5$ fm), 
they pass from being extremely repulsive to reach their maximum attraction.
On the other hand, note that due to the phase space ${\bf P}(r_{12})$ 
is suppressed at small values of $r_{12}$ by at least a factor
$r_{12}^2$. Thus the HF two body probability ${\bf P}(r_{12})$ takes
very small values in the region where the potential is repulsive but
it also misses the maximum attraction region of the $\LL$ interaction. 
The inclusion of a
dependence on $r_{12}$ in the variational wave function allows both to
keep small (and even smaller than in the HF case) the two body
probability in the hard-core region and to increase this two
body probability in the adjacent region where the potential 
reaches its maximum attraction. The steeper $\LL$ potential, the
bigger the difference between the VAR and HF predictions
becomes. Thus, the effects of including $r_{12}-$correlations are much
more pronounced for values of the ratio $\gwl/\gwn$  bigger than 
the standard 2/3 one used in this discussion. 
Therefore, the VAR approach, including correlations,
provides the same binding energies than the HF approach with
significantly less attractive $\LL$ interactions, as we showed in
Figs.~\ref{fig:hfvarll} and~\ref{fig:HF-var}.

\begin{table}
\begin{center}
\begin{tabular}{cccccccc}\hline\tstrut
$l$ & 0 & 1 & 2 & 3 & 4 & 5 & 6 \\\hline\tstrut
${\cal P}_l$ & 0.9727 & 0.02280 & 0.00200 & 0.00075 & 0.00054 &
0.00040  & 0.00028 \\\tstrut
$\sum_{k=0}^{l}{\cal P}_k$ & 0.9727 & 0.9955 & 0.9975 & 0.9982 & 0.9988 
& 0.9991  & 0.9994 \\\hline\tstrut
\end{tabular}
\end{center}
\caption{\small Probabilities ${\cal P}_l$ 
(defined in Eq.~(\protect\ref{eq:mult})) for several waves. Results
were obtained for boron double$-\Lambda$ hypernucleus. A 
local $\LL$ interaction (with parameters 
$\gsl/\protect\sqrt{4\pi}\,=\,2.76$, $\gwl/\gwn\,=\,2/3$ and $\ls\,=1$
GeV, see Table~\protect\ref{tab:var}) 
and  BOY $\Lambda$-core potentials were used. The errors
are always less than one unit in the last digit.  }
\label{tab:mult}
\end{table} 

To finish this section we would like to devote a few words to
the multipolar content of the VAR wave functions. In
Table~\ref{tab:mult} we  show the probability 
${\cal P}_l$, defined in Eq.~(\ref{eq:mult}), 
of finding each of the two $\Lambda$'s with angular momentum $l$ 
 and coupled to $L=0$ for boron. Although a 
particular $\LL$ interaction is used, the 
gross features of the multipolar decomposition 
 do not depend significantly on the selected interaction. 
Only few multipoles contribute\footnote{Note that we show probabilities: the
components of the wave function for each value of $l$  are given, up
to a sign, by the squared root of the numbers presented in the
table.}, being $l=0$
the dominant one  and contributing appreciably only up to the waves  
$l=2$ or $l=3$. The higher the
orbital angular momentum of each $\Lambda$, 
the farer from the origin the corresponding wave function is 
and therefore the smaller the overlap
of the two-body wave function with the attractive part of the
potential is.

Results of Table~\ref{tab:mult}, which show predominance of the 
$l=0$ component in the VAR wave--function, do not contradict the fact 
that correlations are important to lower the energy of 
the double$-\Lambda$ hypernuclei. Indeed, 
the hamiltonian is not diagonal in the basis with well defined
single--particle orbital angular momentum. Hence, non-diagonal matrix
elements contribute significantly to the hamiltonian expectation value.

\subsection{Contribution of the $\phi -$Exchange. }\label{sec:phi}

\begin{table}
\begin{center}
\begin{tabular}{cc|cc|cc}\hline\tstrut
& & \multicolumn{2}{c|}{$\gsl/\sqrt{4\pi}$}& 
\multicolumn{2}{c}{$\gsl/\sqrt{4\pi}$}\\\tstrut
& & \multicolumn{2}{c|}{($\ls = 1$ GeV)}& 
\multicolumn{2}{c}{($\ls = 2$ GeV)}\\\tstrut
 & $\lf$[GeV] & HF & VAR  & HF & VAR \\\hline\tstrut
without $\phi$ & $-$ & $ 3.23  \pm 0.03 $ & $ 2.85  \pm 0.03 $ 
& $2.37 \pm 0.02$ & $2.199 \pm 0.016$ \\\hline\tstrut
            & 1.5  & $ 3.38  \pm 0.03 $ & $ 2.85 \pm 0.03  $ 
& $2.48 \pm 0.02$ & $2.232 \pm 0.018$ \\
with $\phi$ & 2.0 & $ 3.52  \pm 0.03 $ & $ 2.74  \pm 0.03 $  
& $2.58 \pm 0.02$ & $2.16 \pm 0.02$\\
            & 2.5 & $ 3.60  \pm 0.03 $ & $ 2.62  \pm 0.03 $  
& $2.64 \pm 0.02$ & $2.06 \pm 0.02$\\\hline\tstrut
\end{tabular}
\end{center}
\caption{\small Best fit results 
for the $\sigma\LL$-coupling obtained from
different non-local $\LL -$interactions for both HF and VAR
approximations. We compare results for both $\sigma +
\omega$ (already presented in Tabs.~\protect\ref{tab:HF} and
~\protect\ref{tab:var} ) and $\sigma + \omega + \phi -$interactions. 
BOY $\Lambda$-core potentials were used for the three
double$-\Lambda$ hypernuclei. 
The parameter $\gwl/\gwn$ has been fixed
to 2/3. Furthermore, the $\phi -$ couplings have
been fixed, by means of $SU(6)$-symmetry, (see
Eq.~(\protect\ref{eq:phi})). Errors are only statistical.}
\label{tab:phi}
\end{table} 

\begin{figure}
\vspace{-2.cm}
\begin{center}                                                                
\leavevmode
\epsfysize = 750pt
\makebox[0cm]{\epsfbox{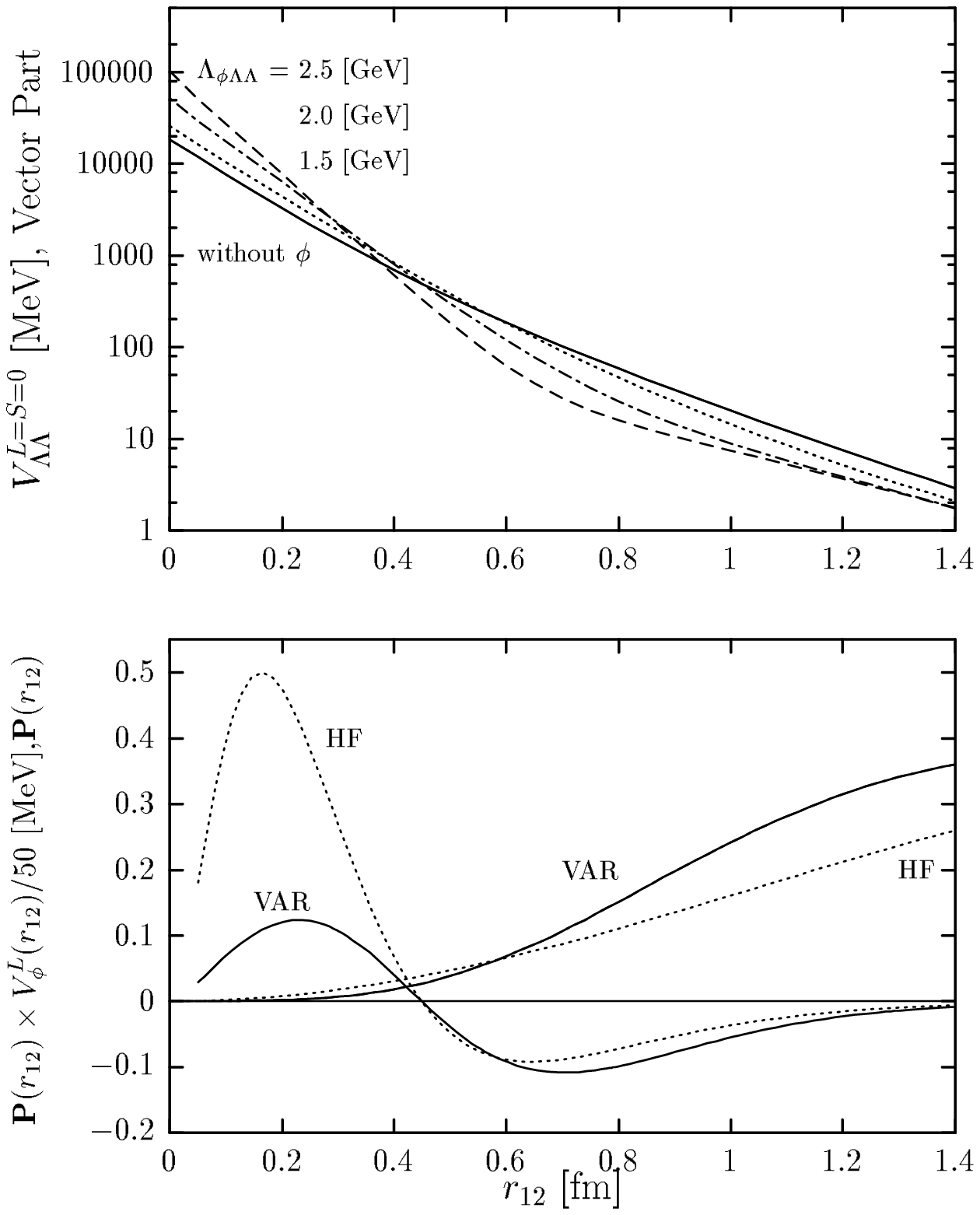}}
\end{center}
\vspace{-10cm}
\caption{\small  Top: $\omega-$ (solid line) 
and $\omega + \phi -$ local pieces of
the $\LL$ potentials predicted by $SU(6)$-symmetry. Three different
values (1.5, 2.0 and 2.5 GeV) for the cutoff $\lf$ have been used. The
larger the $\phi \LL$ cutoff, the bigger is the resulting 
repulsion at the origin. Bottom: Two different set of curves are being
plotted as a function of the $\LL$ relative distance, $r_{12}$ : 
a) HF and VAR probabilities, obtained from the potentials specified in
the first row (``without $\phi$'' and $\ls = 1$ GeV) 
of Table~\protect\ref{tab:phi},  
of finding the two $\Lambda$'s at a relative 
distance $r_{12}$ , b) the product of these probabilities times the
$\lf=2.0$ GeV local $\phi-$potential used in
Table~\protect\ref{tab:phi_1}. This latter set of curves have been
divided by a factor 50 and the area below these curves gives, up to a
factor 50, the expected values given also in
Table~\protect\ref{tab:phi_1}.}
\label{fig:fi}
\end{figure}

In this Subsection we discuss qualitatively 
the effect of the $\phi -$exchange in the $\LL$ potential in the
medium. The $\phi - $potential can be obtained from that of the
$\omega -$exchange given in
Eqs~(\ref{eq:llcomw1})~--~(\ref{eq:ytilda}) 
by the obvious substitutions: $m_\omega \to 
m_\phi$, and $[g,f,\Lambda]_{\omega\LL}  \to 
[g,f,\Lambda]_{\phi\LL} $. In the spirit of the Bonn-J\"ulich models
we use  $SU(6)$ symmetry  to fix the $\phi -$couplings, 
\begin{eqnarray}
g_{\phi \LL} &=& -{\gwl}/{\sqrt2} \nonumber\\ 
f_{\phi \LL} &=&  \sqrt2 \fwl .\label{eq:phi}
\end{eqnarray}
As it was discussed in Subsect.~\ref{sec:ll-int} because within this
model the $\phi$ meson does not couple to nucleons, we do not have much
information about the cutoff-mass $\lf$. Assuming that this cutoff
should be similar to that entering in the $\omega\LL -$coupling and
certainly bigger than the $\phi$ meson mass, we have studied three
different values, 1.5, 2 and 2.5 GeV. Results are presented in
Tables~\ref{tab:phi} and~\ref{tab:phi_1}. We would like to make a few remarks, 
\begin{itemize}

\item The effect of the $\phi-$exchange depends strongly on the value
of the cutoff mass $\lf$, such that this effect passes from being 
small ($\lf = 1.5 $ GeV) to be significant ($\lf = 2.5 $ GeV).
Indeed, this is easily understood by looking at the top graph of
Fig.~\ref{fig:fi} where the $\omega$ and the three 
$\omega + \phi $ potentials considered in Table~\ref{tab:phi} are plotted
(neglecting the small non-local pieces). 
The several $\omega +\phi$ potentials significantly differ in  the
region of relative $\LL$ distances ranging from 0.1  to 1.3 fm, where the 
product of the $\LL$-interaction times the probability (${\bf  P}(r_{12})$)  
of finding the two $\Lambda$ particles at relative distance
$r_{12}$ takes non negligible values. 

As the numbers of Table~\ref{tab:phi_1} indicate, the 
$\phi -$contribution is smaller than the $\omega$ meson one and at maximum it
might be of the order of $20\%$ of the latter one. 
This is not surprising, because as we
mentioned earlier, the exchange of heavier mesons than the $\sigma $
and $\omega$ ones, essentially modify the short range behavior of
the interaction, and hence becomes more significant for processes where
larger momentum transfers than those involved here are relevant. On
the other hand, by the use of different values of the $\sigma -$cutoff
and also different values of $\gwl$ we have already explored a wide
range of short distance behaviors of the effective $\LL -$interaction
in the medium. The above discussion, together with the very scarce set
of data available, have prevented us to include the $\phi -$meson 
contribution  in the present determination of the effective $\LL$
interaction. Its inclusion would require, even assuming
$SU(6)-$symmetry, to deal with at least one more free parameter, and
it might obscure the analysis presented in this work (differences
between HF, perturbative and 
VAR approaches, size of systematic errors $\cdots$). On
the other hand, the possible effect of the $\phi-$exchange potential
in the tail of $\LL$ wave--function, though small, is mostly reabsorbed by the
use of effective $\sigma+\omega -$couplings and thus, the inclusion of the
$\phi$ contribution in the potential would not modify significantly the
mesonic decay of the double$-\Lambda$ hypernuclei studied in 
the next Subsection. However, when
trying to simultaneously analyze the double$-\Lambda$ hypernuclei data
and the $S=-2$ baryon dynamics in the free space, one should not
neglect this contribution because it might lead to sizeable changes in
the values of the $\sigma -$coupling, as the numbers presented in 
Table~\ref{tab:phi} clearly show.

\item A distinctive feature of the results presented in
Tables~\ref{tab:phi} and~\ref{tab:phi_1}, 
is that the character of the $\phi -$ contribution 
though attractive within the VAR scheme,  is repulsive within the
HF one. Hence, the differences between the HF and VAR approaches get
amplified.  $\phi -$exchange does not lead to a purely repulsive
potential, as it 
is the case for the $\omega -$ meson, because the ratio $f/g$ is
positive, instead of negative, and  bigger (in absolute value) 
for the $\phi -$potential 
than for the $\omega-$ one. Thus, the $\phi -$ potential though
much more repulsive than the $\omega - $ one at short distances, 
becomes however attractive at distances bigger than
let us say 0.4 fm  (see Fig.~\ref{fig:fi}
\footnote{Note that in the limit $\lf \to \infty$, the repulsive core
becomes a $\delta$ peak at the origin.}). In the case of the HF
approach, though the
probability {\bf P}$(r_{12})$ is quite small at short distances,
because the huge and repulsive 
values taken by the $\phi -$potential close to zero,
the balance between the repulsive and attractive contributions favors
the former ones. The inclusion of $r_{12}-$correlations in the VAR
scheme allows to have simultaneously 
smaller and bigger values of {\bf P}$(r_{12})$
than in the HF case in the regions where the $\phi -$potential 
is repulsive and attractive respectively. As a net effect, the balance
between the repulsive and attractive contributions favors now
the latter ones. This is easy to appreciate in Fig.~\ref{fig:fi}.

\end{itemize}

\begin{table}
\begin{center}
\begin{tabular}{ccc}\hline\tstrut
& \multicolumn{2}{c}{\phantom{22}
$\left \langle V^L_\phi \right 
\rangle_{\bf\Phi^{(\sigma+\omega)}_{\LL}}$ [MeV]} \\\tstrut
  $\lf$[GeV] & HF & VAR \\\hline\tstrut
             1.5 & 1.95 & $-0.002$   \\
 2.0 & 3.68 & $-1.14$ \\
  2.5 & 4.53 & $-2.33$ \\\hline\tstrut
\end{tabular}
\end{center}
\caption{\small Expected values of the local part of the 
$\phi-$potential, for
several $\phi\LL$ cutoffs, in the $\LL$ state corresponding to the 
HF and VAR solutions (with no $\phi$ contribution and $\ls = 1$ GeV) 
given in the first row of
Table~\protect\ref{tab:phi}. $\phi -$ couplings have
been fixed by means of $SU(6)$-symmetry. For comparison, the expected
values of $\left \langle
V^L_\omega \right 
\rangle_{\bf\Phi^{(\sigma+\omega)}_{\LL}}$ are 14.68 and 10.02
MeV for the HF and VAR wave--functions respectively.   }
\label{tab:phi_1}
\end{table} 

\subsection{Mesonic Decay and Binding Energies of Double$-\Lambda$ 
Hypernuclei}

One might think that the mesonic decay of these double$-\Lambda$ hypernuclei
could depend significantly on the details of the effective 
$\LL$ interaction and thus it could be used to differentiate
between the different potentials shown in
Fig.~\ref{fig:hfvarll}. There are some theoretical
uncertainties\footnote{As can be seen in Fig.~\ref{fig:lcore} and in
Table~\ref{tab:vlcore}, different $\Lambda$-core potentials for helium
lead to significantly different mean squared radius of the $\Lambda$
orbiting around the nuclear core. The mesonic decay process is
quite sensitive to the tail of the $\Lambda$ wave--function and then it
depends strongly on the details of the $\Lambda$-core potential used.}
in the calculation of the mesonic decay of
$^{5}_{\Lambda}$He~\cite{Mo91,St93,OR97}, related to the nuclear 
core--$\Lambda$
interaction, and this has prevented us from 
looking at $^{\phantom{6}6}_{\LL}$He
to check the dependence of the mesonic decay on the details of the 
$\LL$ interaction\footnote{In Refs.~\cite{Mo91} and \cite{Ya97}
the mesonic decay of the double$-\Lambda$ hypernucleus 
$^{\phantom{6}6}_{\LL}$He has been calculated.}. Thus, we have looked at the
case of $^{13}_{\Lambda\Lambda}$B. For this hypernucleus, 
we find that the pionic decay width of $^{12}_{\phantom{1}\Lambda}$B 
changes only around  a 5\% when different $\Lambda$-core 
potentials (BOY, SW1, SW2) are considered.

The mesonic decay has been
computed following the method exposed in Sect~\ref{sec:mes}. We use
the ten VAR non-local potentials presented in
Table~\ref{tab:var}, and we find that the mesonic decay width 
varies by $5\%$ at most, making then this quantity 
unappropriated to choose between the different potentials discussed
above. However, this fact allows us to predict 
the mesonic decay and, by using the potential of
Eq.~(\ref{eq:pot-noso}) in addition to 
Cohen--Kurath spectroscopic factors for the $1p-$shell 
to describe the $^{11}$B, $^{12}$C$^*$  and $^{12}$B$^*$ nuclear
cores, we find 

\begin{eqnarray}
\frac{\Gamma(^{13}_{\Lambda\Lambda}{\rm B} \to X + \Lambda + \pi^0
)}{\Gamma_\Lambda} &=& 0.062 \pm 0.002 \pm 0.002, \label{eq:mes1}\\
&&\nonumber\\
\frac{\Gamma(^{13}_{\Lambda\Lambda}{\rm B} \to X + \Lambda + \pi^-
)}{\Gamma_\Lambda} &=& 0.270 \pm 0.008 \pm 0.008, \label{eq:mes4}
\end{eqnarray}
\noindent where $\Gamma_\Lambda = \Gamma^{\,(p)}_{free} + 
\Gamma^{\,(n)}_{free}$ is the total decay width of the
$\Lambda$ in the vacuum, with  $\Gamma^{\,(\alpha)}_{free}$ 
defined in Eq.~(\ref{eq:free-decay}). 
The central values are obtained with $\gwl/g_{\omega NN} = 2/3$ and
$\ls = 1$ GeV. 
The first set of errors accounts for the statistical error of $\gsl$
whereas the  second one accounts for the spreading of the results 
obtained with the different combinations of ratios $\gwl/g_{\omega NN}$
(1/3, 1/2, 2/3, 1, 4/3)  and values (1 and 2 GeV) for the cutoff
$\ls$.

Further systematic  errors should be added 
to those quoted in Eqs.~(\ref{eq:mes1})~--~(\ref{eq:mes4}) due to both
the uncertainties in the $\Lambda$-core potential (error of about 5\%, as
mentioned above) and the no inclusion of the nuclear core distortion effects 
(errors of about of the same size than the statistical
ones, as discussed in Subsects.~\ref{sec:HF}
and~\ref{sec:VAR}). Adding all the systematic and statistical 
errors in quadratures we end up with total errors of about 7\% 
for both the $\pi^0$ and $\pi^-$ decay widths.

The main ($\geq 95\%$) contribution to these decay widths comes 
from processes where the outgoing nucleon is in a bound state of the
daughter hypernucleus, processes described in 
Eqs.~(\ref{eq:decay1})~--~(\ref{eq:decay2}).

Note that the final nuclear state is described by means
of a simple shell--model supplemented by 
an effective interaction (Cohen-Kurath) for the $1p-$shell
nucleons. Results presented above might depend on the specific details
of the used central and residual interactions. This is a problem,
which is not specific of the mesonic decay of double$-\Lambda$ 
hypernuclei and it is already present in the studies of the 
mesonic decay of single $\Lambda -$hypernuclei. Indeed, if the residual
interaction is neglected, we obtain with the potential of 
Eq.~(\ref{eq:pot-noso}) values for decay widths (in units of 
$\Gamma_\Lambda$)  of $0.074 \pm 0.004 \pm 0.004$ and  
$0.370 \pm 0.008 \pm 0.008$ for $\pi^0$ and $\pi^-$ decays. 
The difference between these values and
those quoted in Eqs.~(\ref{eq:mes1})~--~(\ref{eq:mes4}) might give us
an estimate of the size of our uncertainties. A recent calculation of
the $\pi^-$ decay~\cite{Mo98}, 
where the nuclear states are been constructed in the shell model
within $p-$shell configurations with Cohen-Kurath
interaction~\cite{Co65}, gives $0.325$. This value fits within our range of
results.

To minimize
the effects of this new source of systematic errors, it is interesting to
define the ratios

\begin{eqnarray}
\frac{\Gamma(^{13}_{\Lambda\Lambda}{\rm B} \to X + \Lambda + \pi^0
)}{2\Gamma(^{12}_{\phantom{1}\Lambda}{\rm B} \to X + \pi^0)} &=& 0.63 \pm 0.02
\pm 0.02 \pm 0.02, 
\label{eq:mes3}  \\
\frac{\Gamma(^{13}_{\Lambda\Lambda}{\rm B} \to X + \Lambda + \pi^-
)}{2\Gamma(^{12}_{\phantom{1}\Lambda}{\rm B} \to X + \pi^-)} &=& 0.67\pm
0.02 \pm 0.02 \pm 0.02, \label{eq:mes2}
\end{eqnarray}

\noindent which differ from the naively expected value 1. These
values have been obtained by using the potential of
Eq.~(\ref{eq:pot-noso}) and Cohen-Kurath $1p-$shell spectroscopic factors
to describe the  nuclear cores involved. 
Neglecting the residual interaction the
potential of Eq.~(\ref{eq:pot-noso}) 
leads to similar values (0.65  and
0.72  for $\pi^0$ and $\pi^-$ decays respectively) for these ratios,
as we expected. The meaning
of the first two set of errors in the equations above
 is the same as in Eqs.~(\ref{eq:mes1})~--~(\ref{eq:mes4}). The last
set of error accounts for the systematics due to the no inclusion of 
the nuclear core distortion effects. On the other hand, 
in these ratios, the uncertainties due to the details of the 
$\Lambda$-core potential mostly cancel out.

\begin{figure}
\vspace{-4.5cm}
\begin{center}                                                                
\leavevmode
\epsfysize = 750pt
\makebox[0cm]{\epsfbox{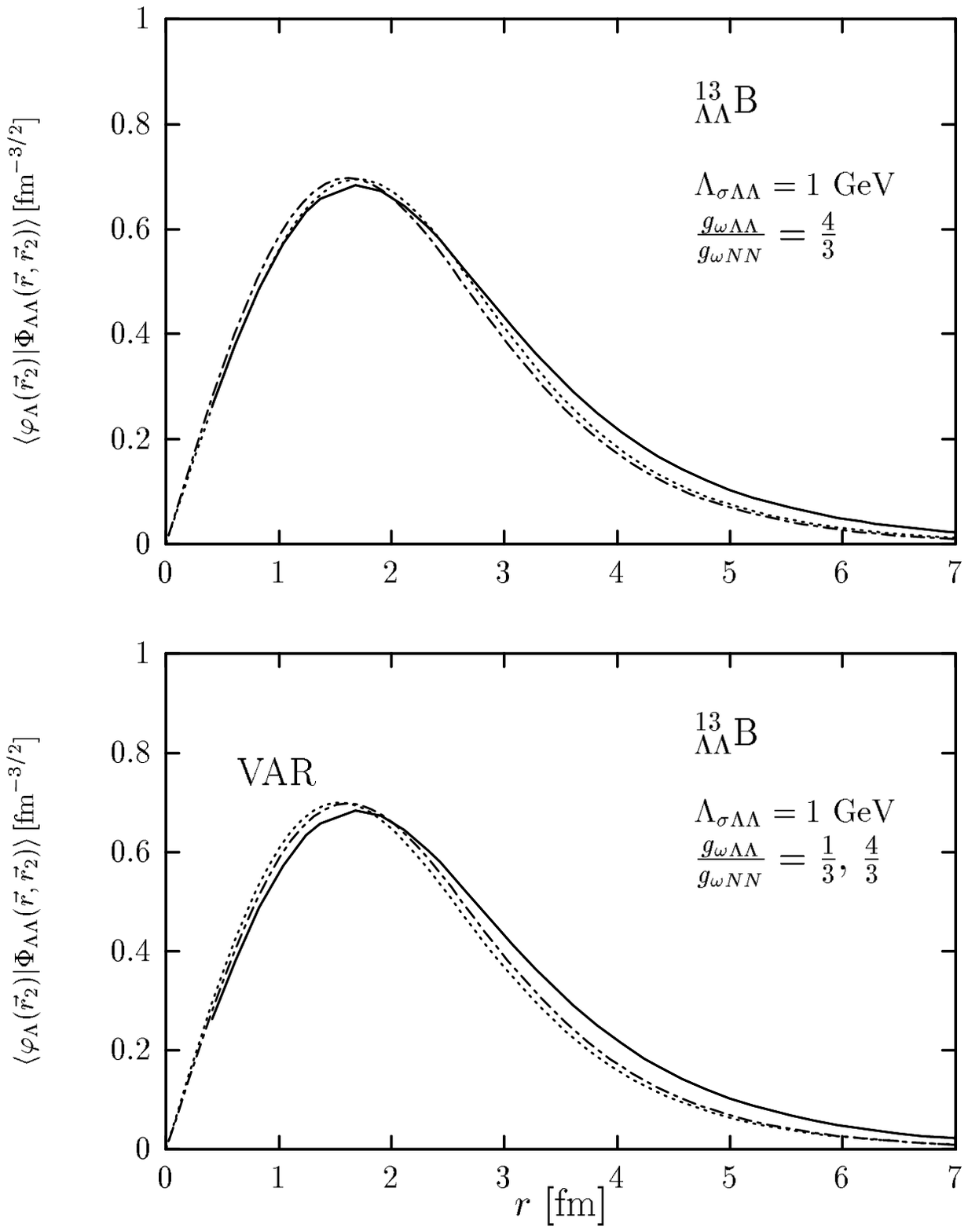}}
\end{center}
\vspace{-11cm}
\caption{\small  The projection  of the two-particle $\Lambda\Lambda$ state
in a double$-\Lambda$ hypernucleus, ${\rm \Phi_{\Lambda\Lambda}}$, 
over the one-particle $\Lambda$ state, $\varphi_\Lambda$,  in the 
corresponding single--$\Lambda$ hypernucleus is shown versus $r$ for
Boron. 
This projection is the overlap defined in Eq.~(\ref{eq:overl}). 
In all cases, the $\Lambda$-core interaction is of the BOY type and
the cutoff parameter $\ls$  is 1 GeV. 
The different lines correspond to different calculations of ${\rm
\Phi_{\Lambda\Lambda}}$ as follows. 
Top: Perturbative (solid), non-local HF (dotted) and VAR 
(dot-dashed) results, with $\gwl/\gwn$ = 4/3 for the two last ones.   
Bottom: Perturbative (solid) and non-local VAR with two values for the
ratio $\gwl/\gwn$,  1/3 (dotted) and  4/3 (dot-dashed). In the VAR and HF
approaches $\gsl$ is obtained from the overall fit to the $B_{\LL}$
data. Within the perturbative scheme, the overlap $\left  \langle
\varphi_\Lambda(\vec{r}_2) | {\rm
\Phi_{\Lambda\Lambda}}(\vec{r},\vec{r}_2) \right \rangle $ coincides
with  $ \varphi_\Lambda(\vec{r})$ ($\Lambda$ wave function in the 
($ ^{A+1}_{\protect\phantom{+1}\Lambda} $$Z$) hypernucleus). }
\label{fig:overl}
\end{figure}

The mesonic decay depends both on the momentum carried by the
outgoing nucleon (the greater the nucleon momentum, the less
effective the Pauli suppression is) and on the tail of the overlap function
defined in Eq.~(\ref{eq:overl}). The first factor is totally
determined by the energy balance in the reaction, which is the same
for all potentials used. Thus, the fact that the 
mesonic widths calculated in the VAR scheme do not
depend appreciably on the specific potential means
that all overlap functions have a similar behavior 
at large distances. This is shown for two potentials in
the bottom plot of Fig.~\ref{fig:overl}. 

On the other hand, HF wave functions lead to results similar, within a
5\%, to those presented in Eqs.~(\ref{eq:mes1})~--~(\ref{eq:mes2}) 
and one would again expect that
both HF and VAR overlap functions have similar tails. 
This can be seen in the top plot of Fig.~\ref{fig:overl}. 
Finally, in this figure we also compare the HF and VAR overlap
functions with the  $\Lambda$ wave function 
in the daughter hypernucleus  ($\varphi_{\Lambda}$). Because the ratio 
$B_{\LL}/2B_\Lambda $ is greater than one, the 
exponential decay at large distances 
of $\varphi_{\Lambda}$ is less pronounced. This, together with the change
in the energy balance of the reaction, which makes more effective the
Pauli blocking for double- than for single--$\Lambda$ hypernuclei, 
explains why the ratios defined in Eqs~(\ref{eq:mes3})~--~(\ref{eq:mes2})
are smaller than one.

\begin{figure}
\vspace{-4.cm}
\begin{center}                                                                
\leavevmode
\epsfysize = 750pt
\makebox[0cm]{\epsfbox{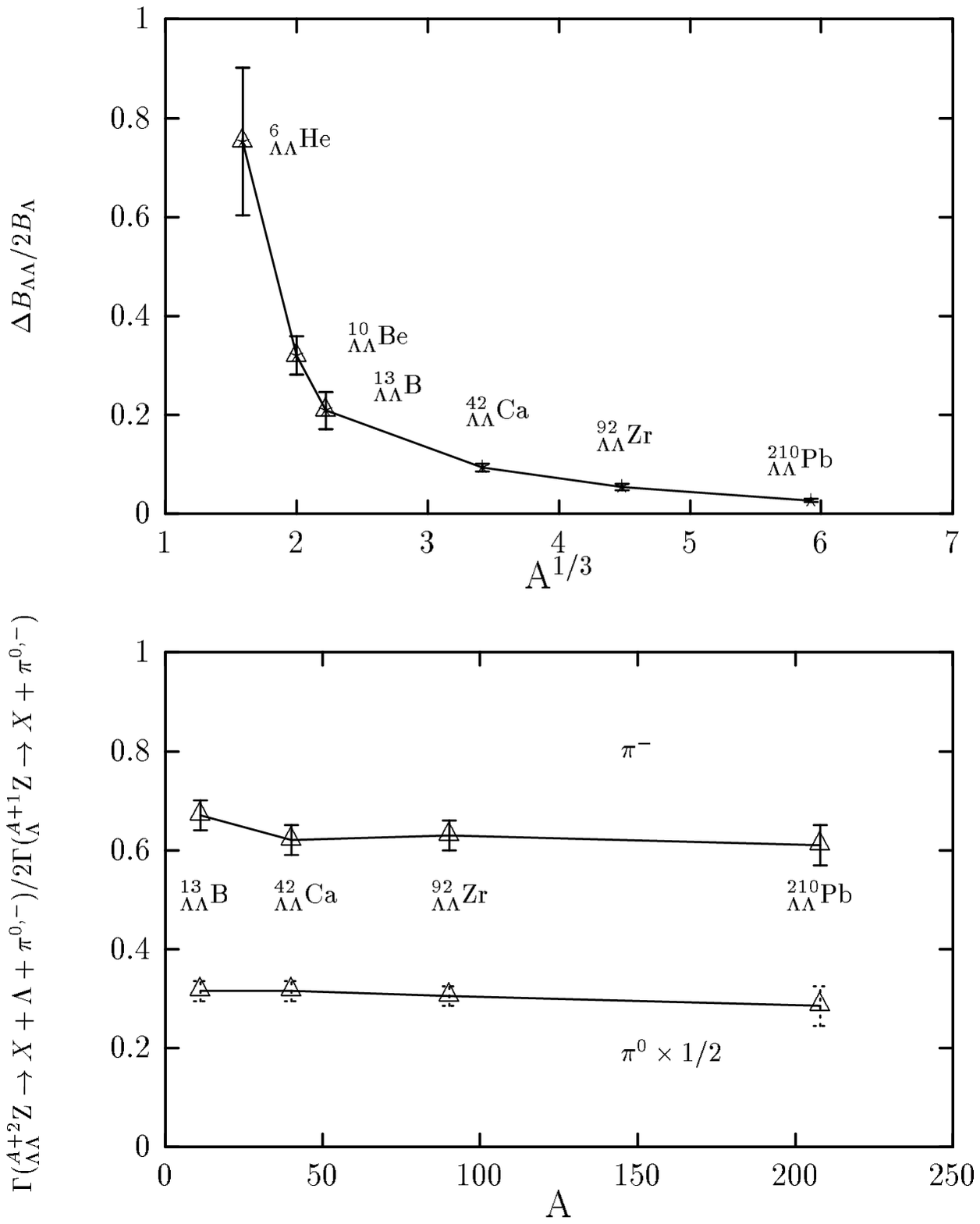}}
\end{center}
\vspace{-11cm}
\caption{\small Ratios $\Delta B_{\LL} / 2 B_{\Lambda}$ (top) and 
$\Gamma(^{A+2}_{\Lambda\Lambda}{\rm Z} \to X + \Lambda + \pi^{0,-} 
)/ 2\Gamma(^{A+1}_{\protect\phantom{1}\Lambda}{\rm Z} \to X + \pi^{0,-})$ 
(bottom) computed for
different double$-\Lambda$ hypernuclei. In the case of the ratio of mesonic
widths with a $\pi^0$ in the final state, the results have been
multiplied by a factor 1/2. BOY  $\Lambda$-core potentials have been
used for all hypernuclei and the
non-local $\LL$ interaction associated to 
$g_{\omega\Lambda\Lambda}= 2g_{\omega NN}/3$ and $\ls$ = 1 GeV has been
used to determine the central value (see Table~\protect\ref{tab:var}). 
Statistical and systematic errors
in the fitted parameter $\gsl$ lead to the error bars in the 
shown ratios, as it is explained in the text after
Eqs.~(\protect\ref{eq:mes1}) - (\protect\ref{eq:mes4}). Experimental
errors in   $B_{\Lambda}$ have not been considered.
In some cases  the size of the errors (systematic and statistical
errors are added in quadrature) is smaller than the symbols.}
\label{fig:a-dependence}
\end{figure}

To finish this section, in Fig.~\ref{fig:a-dependence} we show the
$A$ dependence of the ratios $\Delta B_{\LL} / 2 B_{\Lambda}$ and 
$\Gamma(^{A+2}_{\Lambda\Lambda}{\rm Z} \to X + \Lambda + \pi^{0,-} 
)/ 2\Gamma(^{A+1}_{\phantom{1}\Lambda}{\rm Z} \to X + \pi^{0,-})$. As can be
seen in the figure, we are able to give accurate predictions for
both ratios which also turn out to be  rather independent of the
details, $\gwl$ and $\ls$, of the effective $\LL$  interaction. 

In Ref.~\cite{La97} the $A$-dependece of $\Delta B_{\LL}$ has 
been also studied within a Skyrme-Hartree-Fock approach. There, it is also
found that $\Delta B_{\LL}$ is a decreasing function of $A$. 
We find values for $\Delta B_{\LL}$ within
the broad range of possible results quoted in that reference.

\subsection{Nuclear Medium  
and Free Space $\LL$ Interactions.}\label{sec:medium}

\begin{figure}
\vspace{-7cm}
\hbox to\hsize{\hfill\epsfxsize=0.75\hsize
\epsffile[52 35 513 507]{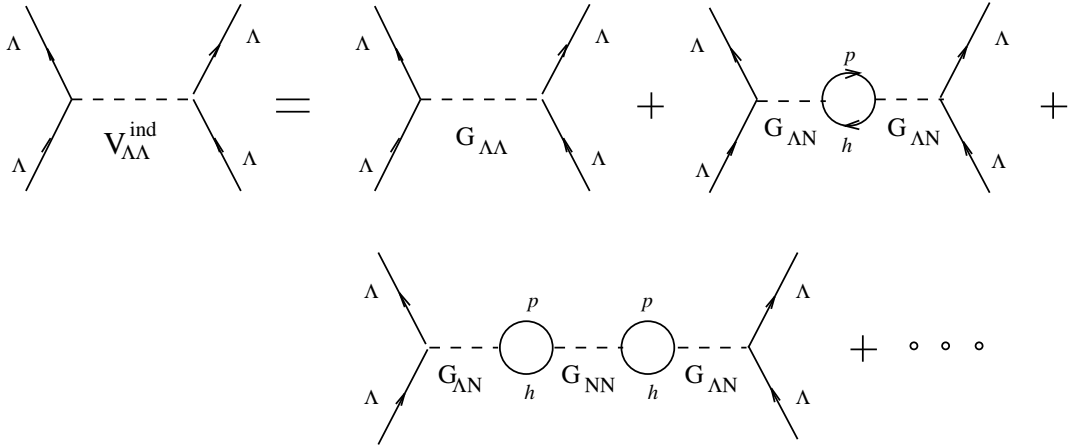}\hfill}
\vspace{1.5cm}
\caption{\small Feynman diagrams relating the $\LL\to\LL,\Lambda N \to
\Lambda N$ and $NN \to NN$\, $G$--matrices
 to the induced interaction ($V_{\LL}^{ind}$) in the nuclear medium,
which approximately accounts for the effective interaction determined
in this work.}
\label{fig:medium}
\end{figure}

It is essential to state that the $\LL$ potentials determined in
this work, effectively describe the dynamics of the $\LL$ pair
 in the nuclear medium, but they do not describe their dynamics in the free
space. Indeed, what has been determined in this work is an effective
interaction in the medium. This effective interaction is usually 
approximated by an
induced interaction~\cite{FW71,Br72,Os90} ($V_{\LL}^{ind}$) 
which is built up in
terms of the $\LL \to \LL$ ($G_{\LL}$), $\Lambda N \to \Lambda N $
($G_{\Lambda N}$)  and $NN \to NN$ ($G_{NN}$)\,\,\, $G-$matrices,
as depicted in Fig.~\ref{fig:medium}. The induced 
interaction, $V_{\LL}^{ind}$, has the virtue of combining the 
dynamics at short distances (accounted by the effective interaction $G_{\LL}$) 
and at long distances (polarization
phenomenon) which is taken care of by means of the iteration of the 
particle-hole ($ph$) excitations (RPA series) through the effective interactions $G_{\Lambda
N}$ and $G_{NN}$. The $G_{\LL}-$, $G_{\Lambda N}-$ and the $G_{N
N}-$matrices can be 
obtained from
the $S=-2$, $S=-1$ and $S=0$ baryon-baryon interactions in the free
space~\cite{YB85,Ya91,Br72}. In the $S=-2$ channel and near threshold, as 
discussed in the introduction, 
one needs to solve the $\LL-\Xi N$ coupled channel $G$--matrix
equations\footnote{The $\LL-\Xi N$ 
mass difference is about 25 MeV. The mass difference
between the $\LL$ and $\Sigma\Sigma$ pairs is about 150 MeV. Thus
one can safely neglect the influence of the latter channel in the
study of the $\LL$ dynamics at threshold. }. In a nuclear medium and 
 with total energies of the order of  $2m_\Lambda$, the $\LL-\Xi N$
coupling might be suppresed due to Pauli blocking 
and thus the data of double$-\Lambda$ hypernuclei would probe primarily the
$\LL$ diagonal element\footnote{It is to say that the contribution of the  
$\LL \to \Xi N \to \LL$ transition would be much smaller than the direct 
$\LL \to \LL $ one (with no $\Xi N$ intermediate states and accounted
for by the $V_{\LL}^{free}$ piece of the $S=-2$ potential), in this context.},
$V_{\LL}^{free}$, of the $\LL-\Xi N$ potential.  Thus, 
$G_{\LL}$ might be roughly approximated by $V_{\LL}^{free}$.

To establish the connection between $V_{\LL}^{ind}$ and
$V_{\LL}^{free}$ is the aim of a future work \cite{Ca98} and it
is out of the scope of this work. As discussed above, it requires to
understand the renormalization of the $\sigma,\omega$ propagation 
(it will be necessary to take into account that both
carriers can excite $ph$ components on their propagation
through the nucleus~\cite{FW71,NOG93}) and
the role played by the $\Xi N$ intermediate states in the nuclear
medium. Note that, within the Bonn-J\"ulich model, the $\phi-$ meson
does not couple to nucleons, and hence its propagation in the nuclear 
medium is not renormalized.

In Table~\ref{tab:bound} we present the smallest values of the $\gsl$
parameter which lead to $^1 S_0$ $\LL$ bound states in the free space,
when  $\sigma +
\omega + \phi-$ $\LL$ interactions, with different cutoffs for the $\phi$
potential, are considered. For the sake of simplicity, we have fixed
$\gwl/\gwn = \frac23$ and $\ls = 1~$GeV. $\phi-$couplings are given in
Eq.~(\ref{eq:phi}). As  can be deduced from the numbers quoted in this
table and those already presented in Table~\ref{tab:phi} the derived HF $\LL$
potentials would lead to $\LL$ bound states. However within the
variational scheme in all cases the potentials are not attractive
enough to bind the two hyperons\footnote{The conclusion is 
the same for the case $\ls = 2~$GeV.}. 
\begin{table}       
\begin{center}
\begin{tabular}{c|c|ccc}\hline\tstrut
& without $\phi$ & \multicolumn{3}{c}{with $\phi$}\\\hline\tstrut
$\lf$[GeV]  & $-$ & 1.5 & 2.0 & 2.5\\\tstrut
$\gsl/\sqrt{4\pi}$ & 3.098 & 3.105 & 3.003 & 2.889 \\\hline
\end{tabular}
\end{center}
\caption{\small Smallest values of the $\gsl-$coupling leading to $^1
S_0$ $\LL$ bound states in the vacuum. Different non-local 
interactions, all of them with $\gwl/\gwn = \frac23$ and $\ls =
1~$GeV, without and with the inclusion of the $\phi -$meson
exchange potential, which coupling to the hyperons is fixed by means of
the $SU(6) -$symmetry,  have been considered. In the latter case three
different cutoffs have been studied.}
\label{tab:bound}
\end{table}

Preliminary results of Ref.~\cite{Ca98} show that the $\LL$
interaction in the medium, $V_{\LL}^{ind}$, 
 is more attractive than that in the free
space, $V_{\LL}^{free}$. Variational results (our best results) 
provide effective interactions unable to bind the $\LL$ pair in the
$^1 S_0$ channel. Therefore we can discard the existence of 
$\LL$ bound states in the free space when only the $\LL$ diagonal element,
$V_{\LL}^{free}$, of the $\LL-\Xi N$ potential is considered. 
An independent confirmation 
of the no existence of $\LL$ bound states is given by the
$\sigma-$coupling ($\gsl/\sqrt 4\pi$ = 2.138) 
extracted from the study of hyperon-nucleon scattering
processes which is well below of any of the $\sigma-$couplings quoted
in Table~\ref{tab:bound}. However, the $\LL-\Xi N$ coupling, though likely 
suppresed in the nuclear medium,
might contribute significantly in the free space. In the model
proposed in Ref.~\cite{Ca97}, the inclusion of the above coupling
leads to more attractive interactions in the $S=-2$ sector, allowing
for values of  the $\LL$  $^1 S_0$ free space scattering length as
large as that of the $nn$ system. Thus, to draw any firm conclusion about the
existence of bound states in the $S=-2$ baryon-baryon sector, 
a combined studied of the $\LL$ and $\Xi N$ systems is needed.

We would like to stress that the important 
issue now is to clarify whether or not the 
difference between the $\gsl/\sqrt 4\pi$ 
values of 2.138 (obtained from hyperon-nucleon scattering data in
Ref.~\cite{Re94}) and of 2.85 (obtained in this work for $\gwl/\gwn =
2/3$  and $\ls = $ 1 GeV within the VAR scheme\footnote{These values
for the ratio $\gwl/\gwn$ and cutoff $\ls$ correspond to those used in
Ref.~\cite{Re94}.}) can be explained in
terms of medium renormalization 
effects and/or the contribution of $\Xi N$ intermediate states 
to the induced 
interaction $V_{\LL}^{ind}$ and/or the contribution of heavier mesons not
included here, such as the $\phi$ meson.
 As discussed in Subsect.~\ref{sec:phi}, the inclusion of the
$\phi-$exchange piece of the $\LL$ potential reduces in some extent 
the difference between the $\gsl$ couplings mentioned above. For
instance, if a $\lf$ cutoff of 2.5 GeV is used, one finds values for 
$\gsl/\sqrt 4\pi$ of the order of 2.6 (Table~\ref{tab:phi}).

Finally a word of caution must be said here. The J\"ulich group has
also shown~\cite{Ho95,Ho96} that the correlated $2\pi$ 
and $K\bar K$ exchanges lead to a value for the ratio $\gsl/\gsn$ of
0.49. This value is much smaller than the results obtained 
in any of the present $YN$  
models~\cite{Ma89,Na75,Ho89,Re94,St88,St90}, for which this ratio
varies from 0.58 to 1. Therefore, the intermediate range attraction in
the $\Lambda N $ and $\Lambda \Lambda$ channels will be significantly
reduced if one assumes the results of the microscopical calculations of
Refs.~\cite{Ho95,Ho96}. Then, a value for the ratio $\gwl/\gwn$
smaller than 2/3, predicted by $SU(3)$ and commonly accepted, would be
needed to reproduce the $\Lambda N$ scattering data. Thus, once all
nuclear medium effects are understood, it will also be worth
studying if a new set of parameters with $\gsl/\gsn = 0.49$
and $\gwl/\gwn < 2/3$ provides a 
simultaneous acceptable description of both double$-\Lambda$ hypernuclei 
and $\Lambda p$ scattering data.

\section{Conclusions}
\label{sec:concl}

The dynamics of the $\LL$ system can not be studied directly in the
free space. $\Lambda$ $p$ scattering  provides an indirect
source of information about it. Thus, the physics 
of double$-\Lambda$ hypernuclei provides an independent and valuable 
method to confirm the findings obtained in the studies of
hyperon-nucleon scattering. From these studies, it is commonly
accepted that there exist three kinds of realistic 
models for the $\LL$ diagonal element of the $\LL-\Xi N$ potential, 
those developped by the  
Nijmegen, T\"ubingen and J\"ulich groups. The J\"ulich OBE potential for the
hyperon-hyperon  interaction is a
natural extension of the Bonn model for the
$NN$-interaction~\cite{Bonn}. In this paper we have 
used for the first time J\"ulich type $\LL$ OBE potentials to study 
the dynamics of  double$-\Lambda$ hypernuclei. We have found:

\begin{itemize}

\item Effective $\LL$ interactions in the medium 
extracted from J\"ulich type models
provide a simultaneous fairly good description of the binding
energies of the three known double$-\Lambda$ hypernuclei.

\item Uncertainties in the exact nature of the $\Lambda$-nuclear core 
dynamics lead to  uncertainties  in the 
determination  of the effective 
$\LL$ potential in the medium smaller than those due to 
the experimental errors in the fitted quantities $\Delta B_{\LL}$. 
Thus, the ambiguities in the
determination of the $\Lambda$-core potential do not constitute an important
obstacle to learn details about the $\LL$ interaction.

\item The HF and the perturbative approaches, 
discussed in Subsects.~\ref{sec:HF}
and \ref{sec:pert} respectivelly, lead to quite similar effective $\LL$
potentials. 

\item The inclusion of $\LL$ correlations in the variational
scheme provides a better understanding of the dynamical features 
of the system. It also enables  
less attractive effective $\LL$  potentials than in the HF or
perturbative approaches and more similar to those extracted from
hyperon-nucleon scattering to describe the
binding energies of the double$-\Lambda$ hypernuclei.

\item Only few waves (up to $l=3$) contribute appreciably 
to the multipolar expansion of the VAR wave functions.

\item The inclusion of the $\phi -$exchange might be relevant to
understand the effective $\LL$ interaction derived in this work in
terms of the free space one.

\item The existing double$-\Lambda$ hypernuclei data can not 
conclusively favour any particular choice neither of the ratio 
$g_{\omega\Lambda\Lambda}/g_{\omega NN}$ in the 
interval $[1/3,4/3]$ around the $SU(3)$ prediction 2/3, nor of the
cuttoff $\ls$ in the range = 1--2 GeV. Thus, we end up with a
whole family of effective 
$\LL$ potentials describing the ground state binding energy
of the three known double$-\Lambda$ hypernuclei.

\item The mesonic decay of $^{13}_{\Lambda\Lambda}{\rm B}$ and
the binding energies and pionic decay widths 
of heavier double$-\Lambda$ hypernuclei (not discovered
yet) turned out to be rather independent of the details of 
the effective $\LL$ interaction within
the family of potentials described in the previous point. This fact has
allowed us to predict accurately, for the very first time, the
mesonic decay widths of medium and heavy double$-\Lambda$ hypernuclei.

\item We have discarded the existence of $^1 S_0$ $\LL$ bound states
in the free space if the  $\LL-\Xi N$ coupling is negligible.

\end{itemize}

The natural continuation of this work~\cite{Ca98} is the study of the
nuclear medium modifications of the $\LL$ interaction. Such study,
together with the effective $\LL$ interactions derived here, will allow
us  to better understand the dynamical features of the $\LL$ interaction,
at low energies, in the vacuum. Thus, it will be possible to contrast 
features of such interaction from two different 
and independent sources of
data: hyperon-nucleon scattering and $\LL$ hypernuclei. Statisticaly
improved and new experimental data on double$-\Lambda$ hypernuclei,
exploring not only light but also medium and heavy nuclei, 
will be extremely valuable to achive such an objective.

\section*{Acknowledgments}
We would like to acknowledge useful discussions with E. Buend\'\i a,
F. G\'alvez, A. Ramos,  L.L. Salcedo and A. Sarsa. J.C. thanks the 
Departamento de F\'\i sica Moderna in Granada for its
kind hospitality during the early stages of this project. 
This research was supported by DGES under contract PB95-1204 and by
the Junta de Andaluc\'\i a. J.C. acknowledges the Spanish
{\rm Direcci\'on General de Ense\~nanza Superior}
(Ministry of Science and Education) for support through
a postdoctoral fellowship.


\appendix
\section*{Appendix \\ Matrix Elements in the VAR Scheme}

The basic integrals to compute the matrix elements of the hamiltonian
of Eq.~(\ref{eq:sch}) in the basis defined in Eq.~(\ref{eq:ans}) are: 
\begin{eqnarray}
\Gamma_{lmn}(a,c) 
&=&8\pi^2\int_0^{ + \infty}dr_1\, r_1^l \int_0^{+ \infty}dr_2\, r_2^m 
\int_{|r_1-r_2|}
^{r_1+r_2}dr_{12}\,r_{12}^n~e^{-\left (
a\left(r_1+r_2\right)+c\,r_{12}\right)}\,,
\end{eqnarray}
with $a,c$  and $l,m,n$  real positive and  integer numbers
respectively. To compute the matrix elements of 
 the $\Lambda$ kinetic terms, of the HE term, 
of the $\Lambda$-core potentials and  of
the local part of the $\LL$ interaction, only
non-negative values of $l,m$ and $n$ are needed. In this case, 
all the integrals can
be obtained from $\Gamma_{000}$  by means of a 
recursion formula proposed in Ref.~\cite{Sa67}:
\begin{eqnarray}
\Gamma_{lmn} &=& \frac{1}{2a}\left [l\,\Gamma_{l-1,m,n} + m\, \Gamma_{l,m-1,n} 
+ B_{lmn} \right],\\
B_{lmn} &=& \frac{1}{a+c}\left [l\, B_{l-1,m,n} + n\, B_{l,m,n-1} 
+ \delta_{l\,0}\frac{(4\pi)^2 (m+n)!}{(a+c)^{m+n+1}} \right],
\end{eqnarray}
with $\Gamma_{000}$ given by:
\begin{eqnarray}
\Gamma_{000} &=&  \frac{(4\pi)^2}{2a(a+c)^2}\,.
\end{eqnarray}
To compute the matrix elements of the non-local part of the $\LL$
interaction, integrals of the type $\Gamma_{l,m,-1}$, with $l,m =
0,1,2 \cdots $ are also needed. These can be obtained from a recursion
as well~\cite{jca97},
\begin{eqnarray}
\Gamma_{l,m,-1} &=& \frac{1}{2a}\left [l\,\Gamma_{l-1,m,-1} 
+ m\, \Gamma_{l,m-1,-1} 
+ \frac{l!\,m!}{l+m+1}\frac{(4\pi)^2}{(a+c)^{l+m+1}} \right],
\end{eqnarray}
with $\Gamma_{0,0,-1}$ given by
\begin{eqnarray}
\Gamma_{0,0,-1} &=&  \frac{(4\pi)^2}{2a(a+c)}\,.
\end{eqnarray}

A final detail, related to the non-local part of the $\LL$ interaction,
concerns to the implementation of $\vec{\nabla}_{12}$ when acting on
functions of $r_1$,$r_2$ and $r_{12}$. The operator
$\vec{\nabla}_{12}$ has to be understood as the gradient on the
direction $\vec{r}_{12} = \vec{r}_1-\vec{r}_2$ when the coordinates
of the center of mass of the 
$\LL$ pair, $\vec{R}_{cm} = (\vec{r}_1+\vec{r}_2)/2$, are kept fixed. Thus, 
\begin{eqnarray}
\vec{\nabla}_{12} f(r_1,r_2,r_{12}) &=& \frac12 \left ( \vec{\nabla}_1 - \vec{\nabla}_2
\right ) f(r_1,r_2,r_{12}), \\
& = & \frac12 \left ( \frac{\vec{r}_1}{r_1}\frac{\partial}{\partial
r_1} - \frac{\vec{r}_2}{r_2}\frac{\partial}{\partial r_2} + 
2 \frac{\vec{r}_{12}}{r_{12}}\frac{\partial}{\partial r_{12}}\right )
f(r_1,r_2,r_{12}),   
\end{eqnarray}
with $\vec{\nabla}_{1\,(2)}$ the gradient on the direction 
$\vec{r}_{1\,(2)}$ when $\vec{r}_{2\,(1)}$ is kept fixed.

\end{document}